\documentclass[12pt]{article}

\usepackage{color}

\usepackage{graphicx}
\usepackage{float}
\usepackage{amsmath}
\usepackage{amsthm}
\usepackage{amssymb}
\usepackage{amscd}
\usepackage{amsfonts}
\usepackage{makeidx}
\usepackage{enumerate}

\usepackage{mathrsfs}
\usepackage{esint}
\usepackage{cancel}

\newtheorem{teo}{Theorem}[section]
    \newtheorem{lem}[teo]{Lemma}
    \newtheorem{prop}[teo]{Proposition}

    \newtheorem{obs}[teo]{Remark}

    \newtheorem*{dem}{\textsc{Proof}}
    \newcommand{\bdem}{\begin{dem}}
    \newcommand{\edem}{\end{dem}}
     \newcommand{\be}{\begin{equation}}
    \newcommand{\ee}{\end{equation}}
     \newcommand{\ba}{\begin{array}}
    \newcommand{\ea}{\end{array}}
\newcommand{\beqn}{\begin{eqnarray}}
    \newcommand{\eeqn}{\end{eqnarray}}
    \newcommand{\bl}{\begin{lem}}
    \newcommand{\el}{\end{lem}}
    \newcommand{\bp}{\begin{prop}}
    \newcommand{\ep}{\end{prop}}
\newcommand{\ds}{\displaystyle}

      \newcommand{\vf}{\varphi}
      \newcommand{\ve}{\varepsilon}

    \newcommand{\R}{\mathbb{R}}
    \newcommand{\C}{\mathbb{C}}
     
     \newcommand{\no}{\noindent}
     \newcommand{\Res}{{\rm Res}}\newcommand{\const}{{\rm const}}
     \newcommand{\Ort}{\operatorname{Ort}}\newcommand{\f}{{\bf f}}

\def\Re {{\rm Re\, }}
 \def\Im {{\rm Im\,}}
\def\Res {{\rm Res\,}}

\begin{document}
\title{Trapped modes and resonances for thin horizontal cylinders in a two-layer fluid}
\author{
\large{ P.  Zhevandrov$^1$},\\
\large{A.  Merzon$^2$},\\
\large{M.I. Romero Rodr\'iguez$^3$}\\
\large{and J.E. De la Paz M\'endez$^4$}\\
{\small{\it $^1$ Facultad de  Ciencias F\'\i sico-Matem\'aticas, Universidad Michoac{a}na}},\\[-2mm]
{\small  Morelia, Michoac\'{a}n, M\'{e}xico}\\[-2mm]
{\small{\it $^2$ Instituto de F\'\i sica y  Matem\'aticas, Universidad Michoac{a}na}},\\[-2mm]
{\small  Morelia, Michoac\'{a}n, M\'{e}xico}\\[-2mm]
{\small{\it $^3$ Facultad de Ciencias B\'asicas y Aplicadas, Universidad Militar Nueva Granada}},\\[-2mm]
{\small{Bogot\'a, Colombia}},\\[-2mm]
{\small{\it $^4$ Facultad de Matem\'aticas II, Universidad Aut\'onoma de Guerrero}},\\[-2mm]
{\small{Cd. Altamirano, Guerrero, M\'exico}}\\[-2mm]
{\small{\it E-mails}: pzhevand@gmail.com,}
{\small anatolimx@gmail.com,}\\[-2mm] {\small maria.romeror@unimilitar.edu.co,} {\small jeligio12@gmail.com}}
\maketitle


\begin{abstract}
  Exact solutions of the linear water-wave problem describing oblique  waves over a submerged horizontal cylinder of small (but otherwise fairly arbitrary) cross-section in a two-layer fluid are constructed in the form of convergent series in powers of the small parameter characterizing the ``thinness" of the cylinder. The terms of these series are expressed through the solution of the exterior Neumann problem  for the Laplace equation describing the flow of unbounded fluid past the cylinder. The solutions obtained describe trapped modes corresponding to discrete eigenvalues of the problem (lying close to the cut-off frequency of the continuous spectrum) and resonances lying close to the embedded cut-off. We present certain conditions for the submergence of the cylinder in the upper layer when these resonances convert into previously unobserved embedded trapped modes.
\end{abstract}

\section{Introduction}

The study of trapped modes (finite energy solutions) for water waves in infinite channels which appear under perturbations dates essentially from  Ursell's paper \cite{Ursell} published in 1951. He proved that a submerged thin circular cylinder (which is a perturbation in this case) gives rise to the appearance of finite-energy solutions with an eigenvalue below the cut-off. Since that date, large amount of new results was obtained (see \cite{Kuz-Maz-Va, Naz2009} and references therein), especially with respect to the existence of trapped modes. In the majority of those investigations, no use of asymptotics was made and the problem  did not depend on the smallness of the obstruction. Of course, in the latter case, it is impossible to obtain explicit formulas for the eigenvalues, although the problem was extensively studied by means of numerical methods in order to obtain quantitative information about them (see, e.g., \cite{McIver-Linton} and references therein). On the other hand, when the obstruction is in some sense small, one can use asymptotic methods in order to construct explicit (albeit approximate) formulas for the solution. For example, in \cite{McIver} cylinders of arbitrary  cross-section were considered and  approximate solutions were constructed  under the assumption that the cross-section is symmetric with respect to a   vertical plane passing through the axis of the cylinder and its area is small. The asymptotics in \cite{McIver} was constructed by means of the matching technique \cite{Ilin}. A successful use of matching in this case is not at all surprising, the problem being an elliptic equation in a half-plane with a small orifice (a number of problems of this kind is considered in the book \cite{Ilin}). Another example is a small protrusion  on the bottom \cite{Kuz, MIRZh}; in the last paper an exact solution was obtained by means of the reduction of the problem to integral equations (see also \cite{Lint}).

  In paper \cite{GarZhev}  the problem of waves trapped by a submerged thin cylinder with  fairly arbitrary cross-section in a homogeneous fluid without any symmetry conditions was considered.  This resulted in finding exact solutions in the form of series in powers of $\ve$ and $ \ve \ln \ve$, where $\ve$ characterizes the ``thinness" of the cylinder, by means of a technique similar to \cite{ MIRZh}, \cite{Mar}--\cite{19}. The leading term of the series coincided, of course, with the result of \cite{McIver} in the case of symmetric cylinders. The goal of the present paper is to extend these results to the case of a two-layer fluid.

   Modes trapped by submerged cylinders in a two-layer fluid were studied numerically, for example, in  \cite{LintCad, Saha} and analytically in \cite{NazVid2009, NazVid};  papers \cite{Saha, NazVid} contain a rather complete bibliography.  In the case of two superimposed layers there can exist two propagating modes because of the presence of the free surface and the interface. The continuous spectrum occupies the ray $\lambda\geq \Lambda_1$ ($\lambda=\omega^2/g$ is the spectral parameter, $\omega$ is the frequency and $g$ is the acceleration of gravity) and there is an embedded cut-off $\Lambda_2>\Lambda_1$ such that the multiplicity is 2 for $\Lambda_1<\lambda<\Lambda_2$ and 4 for $\lambda>\Lambda_2$ (the values of the cut-offs $\Lambda_{1,2}$ are expressed through the density ratio of the layers, the depth of the upper layer, and the wavenumber along the cylinder). This problem arose in part as an attempt to understand the results of \cite{Kuz1993} (see also \cite{KuzMcI}). In that paper, two types of trapped modes in a two-layer fluid spanned by a horizontal cylinder were obtained, the first with frequency lying outside the continuous spectrum and the second embedded in it (the interest in this problem was stimulated by the proposed construction of tube bridges in Norwegian fjords which have a manifestly two-layer structure). The numerical results of \cite{LintCad} suggested that only the first type of trapped modes can exist and the second type always presents a leaky behavior. However, when one cylinder was changed to two cylinders, an embedded trapped mode appeared to the  left of $\Lambda_2$ for certain values of the distance between the centers of the cylinders. The computations in \cite{LintCad} were based on the observation made in an earlier paper \cite{LinCadScat}; in that paper the phenomenon of total reflection by one submerged cylinder was observed at particular frequencies.  The existence of the discrete eigenvalue (trapped mode with frequency outside the continuous spectrum) was proven in \cite{NazVid2009}, and in \cite{NazVid} the asymptotics of the discrete eigenvalue was obtained for almost equal densities of the layers or for the upper layer of vanishing density (both limits, as shown in \cite{NazVid}, are singular). Nevertheless, explicit formulas for the frequencies of trapped modes and analytical description of  embedded trapped modes are apparently lacking.

 One of our goals here is to provide a theoretical understanding of the numerical results from \cite{LintCad} mentioned above  for a  complete two-layer problem. The technique of \cite{MIRZh, GarZhev, AyaCanoZh, MIRZhJFM}  enables one to construct {\it exact}  solutions in the form of convergent series in powers of the small parameter $\ve$ characterizing the perturbation, thus providing a rigorous justification of the results. The main difference from \cite{MIRZh, GarZhev} consists in the fact that instead of one integral equation the problem reduces to a system of two equations due to the existence of two wave modes. Similar  results were obtained in our earlier paper \cite{MIRZhJFM} in the shallow-water approximation; here we are going to extend some of  those results to the case of full potential problem with submerged cylinders. We note the importance of the fact that our solutions are exact; while for the discrete eigenvalues one can use the well-known tools of the perturbation theory to prove the closeness of approximate eigenvalues to the exact ones, for the embedded eigenvalues these tools are as a rule not sufficient and the corresponding proofs are quite involved (see, e.g., \cite{Naz}). As in \cite{MIRZhJFM}, we understand the term ``exact solutions'' in the sense that we guarantee the existence of convergent series for them and explicitly calculate the leading terms.

 It turns out, as in \cite{MIRZh, GarZhev} that the use of the Fourier transform significantly simplifies the calculations due to the fact that the Fourier transform of the fundamental solution $K_0$ of the modified Helmholtz equation is an elementary function.

 We will construct exact solutions describing trapped modes corresponding to discrete eigenvalues which lie outside the continuous spectrum and close to the first cut-off for one cylinder in the upper or lower layer, and will construct resonances (leaky modes) that are complex and lie close to the second embedded cut-off (see, e.g.,  \cite{Duan} for the definition of resonances). It turns out that for the cylinder in the upper layer these resonances can convert into embedded eigenvalues for a symmetric cylinder whose submergence is related to the hydrodynamic properties of the cross-section of the cylinder and the profile of the propagating mode. This apparent contradiction to the numerical results from \cite{LintCad} (where no embedded eigenvalues for {\it one} submerged cylinder were found numerically) does not in fact contradict anything: first, the cylinder in that paper is not at all thin (its diameter is half the depth of the upper layer), second, the graphs of the reflection and transmission coefficients from \cite{LinCadScat} in the case of the cylinder in the upper layer present characteristic asymmetric form of Fano resonances which as a rule are observed close to an embedded eigenvalue (see \cite{Ship1, Ship2, ZAMP}). For the cylinder in the lower layer, these graphs do not present the Fano ``dips'', which is in total agreement with  our results that show that in the latter case the resonances never convert into embedded trapped modes. Since embedded trapped modes, as is well-known, are highly unstable under perturbations, the fact that they were not observed numerically is not at all surprising.

 Heuristically, the appearance of embedded trapped modes produced by one submerged cylinder in the upper layer can be explained as follows. The presence of a small obstacle is equivalent, in some sense, to the presence of a point source; if this point source is positioned in such a way that it is ``invisible'' (orthogonal) to the propagating mode, then the resonance decomposes into a plane propagating wave and a trapped mode which are independent of each other. This phenomenon is  possible due to the fact that the vertical profile of the propagating mode is not necessarily monotonic in the upper layer. For the cylinder in the lower layer, this profile is monotonic and hence no trapped mode appears.

We emphasize the fact that we are treating a singular perturbation problem (there are no eigenvalues, embedded or not, when $\ve=0$), and, although the solution is expressed in the form  of a series in powers of $\ve$, the coefficients of this series are in their turn functions of $\ve$ (like $\exp\{-\ve|x|\}$, where $x$ is the spatial coordinate) and do not admit a regular expansion in $\ve$ for all (unbounded) $x$. Finally, we note that our results furnish explicit formulas for the frequencies of trapped modes and therefore can be used in direct engineering applications, e.g., to estimate the ``dangerous'' resonant frequencies. 

The paper is organized as follows: in section 2 we give the formulation, present our main results and describe some limiting forms of our formulas for the asymptotic regimes studied in \cite{NazVid}. In section 3 we study the problem of a cylinder in the upper layer, construct the discrete eigenvalue and the resonance which is close to the second cut-off and indicate the conditions when this resonance converts into an embedded eigenvalue. In fact, this last result presented in section 3.6 and Theorem \ref{LAN3} constitutes the most interesting result of the present paper. The first of our results (Theorem \ref{1} dealing with the discrete eigenvalue for the cylinder in the upper layer) was already announced with a sketch of the proof in \cite{MIRZMMAS}; we give a detailed and slightly improved proof in section 3.4. In section 4 we do the same for the cylinder in the lower layer, but prove that the resonance never converts into an eigenvalue. In Appendices 1-3 we reduce the initial problem to systems of integral equations and prove some technical results.

\section{Formulation and main  results}
\setcounter{equation}{0}
\subsection{Formulation}\label{2.1}

\begin{figure}[!h] 
\begin{center}
\includegraphics[width=8cm, height=7cm]{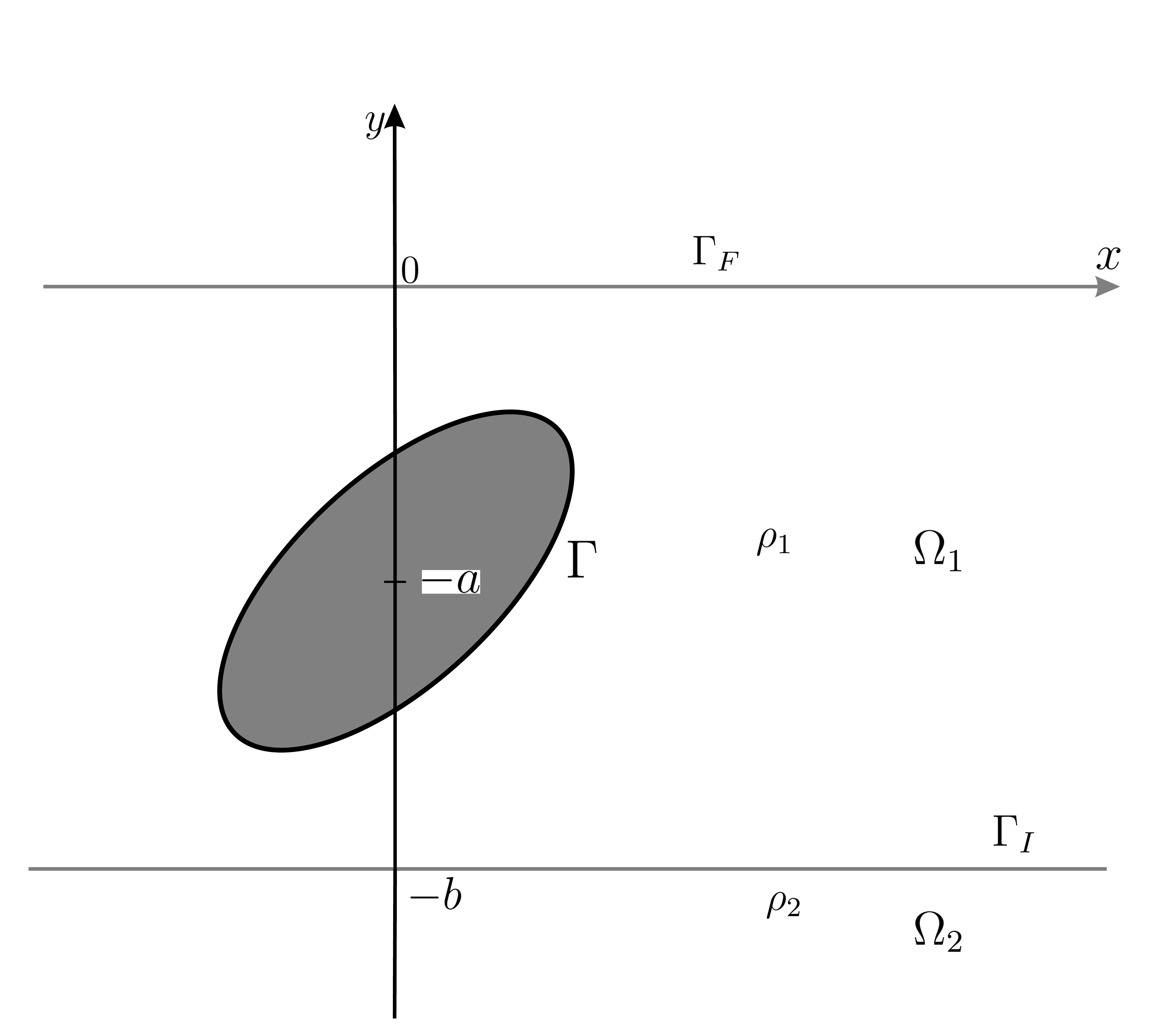}
\end{center}
\caption{Problem $U$}  \label{elipse_inclinada}
\end{figure}

\begin{figure}[!h]
\begin{center}
\includegraphics[width=8cm, height=7cm]{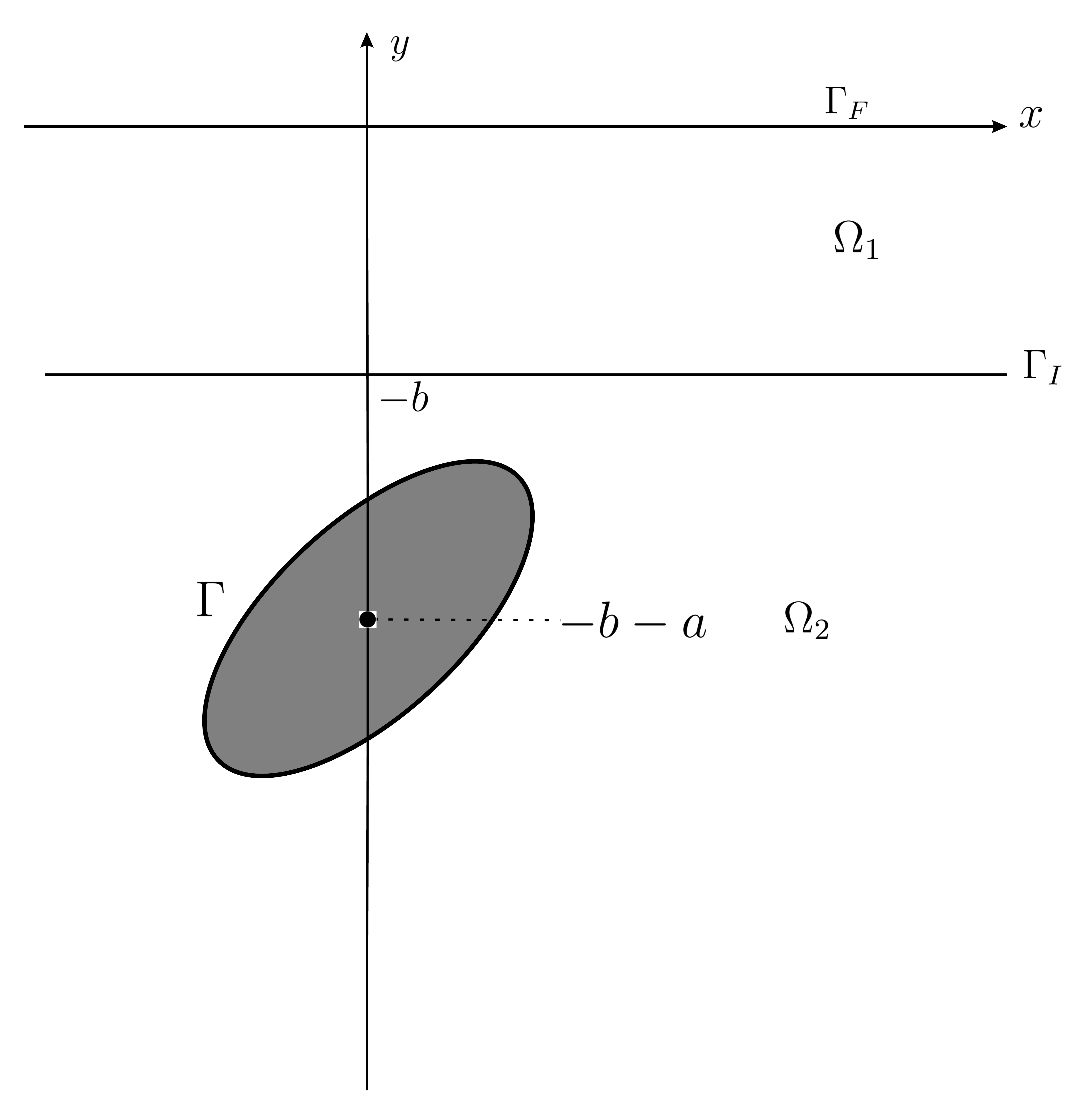}
\end{center}
\caption{Problem $L$}  \label{ANEWFIG}
\end{figure}

Assume that the upper layer of the fluid of density $\rho _1$ is confined to
$ -b < y < 0$, and the  lower layer of density $\rho_2>\rho_1$ occupies the lower half-space $y < - b$. Further, assume that a cylinder is immersed in the upper (problem $U$) or lower  (problem $L$) layer (see Fig.\;\ref{elipse_inclinada} and Fig.\;\ref{ANEWFIG})


We will assume that the cylinder does not intersect the interface $y=-b$, is parallel to the $z$-axis, $x$ being the horizontal coordinate orthogonal to the cylinder. We look for the potentials in the layers in the form ${\rm exp}\{ i(kz-\omega t)\} \phi_{1,2}(x,y)$ where $t$ is time and we assume that $k>0$ (oblique incidence) and $\Re \omega>0$.
Let the surface of the cylinder be defined by
$$
\Gamma =\left\lbrace y = -a + \varepsilon  Y(t), \quad x= \varepsilon  X(t), \quad - \pi \leq t \leq \pi\right\rbrace
$$
for  problem $U$ and
 $$
\Gamma =\left\lbrace y = -b-a + \varepsilon  Y(t), \quad x= \varepsilon  X(t), \quad - \pi \leq t \leq \pi\right\rbrace
$$
for  problem $L$, where
 $
 \dot{X}^2 +\dot{Y}^2 \neq 0,
 $
$X$ and $Y$ are  $2 \pi$-periodic $C^{\infty}$ functions with zero mean,
$
\int_{-\pi}^{\pi} X(t)\, dt = \int_{- \pi}^{\pi} Y(t)\, dt = 0,
$
and $X$ and $Y$ describe a  simple closed curve in the plane $(x, y)$ whose interior stays to the left when $t$ increases. Here $\varepsilon\to0$ is a small parameter and $t$ is a parameter along the curve $\Gamma$ and has nothing to do with the time coordinate.
In both problems we assume that $b>0$ and $a>0$, moreover, for problem $U$ we assume that $a<b$.

The functions $\phi_{1,\,2}$ satisfy the following system:
\begin{align} \label{etiqueta_1}
\Delta \phi_1 - k ^2 \phi_1 &= 0\qquad \mathrm{in} \qquad \Omega_1,
\\ \nonumber\\\label{etiqueta_1_2}
 \Delta \phi_2 - k ^2 \phi_2 &= 0\qquad \mathrm{in} \qquad \Omega_2,
\\ \nonumber\\  \label{etiqueta_2_2}
 \phi_{1\,y} &=\lambda \phi_1 \quad \mathrm{on} \quad \Gamma_F,
\\\nonumber\\ \label{etiqueta_2_3}
\beta (\phi_{1\, y} -\lambda\, \phi_1) &=\phi_{2\, y}-\lambda\, \phi_2 \quad \mathrm{on}\quad \Gamma_I,
\\ \nonumber\\  \label{etiqueta_2_4}
\phi_{1\, y} &=\phi_{2\, y}  \quad \mathrm{on} \quad \Gamma_I.
\end{align}
Here $\beta = {\rho_1}/{\rho_2}<1$, $\Gamma_F=\{y=0 \}$ is the free surface at rest,
$\Gamma_I=\{ y=-b \}$ is the interface at rest, $\Omega_1$ for problem $U$ is the exterior of $\Gamma$ in the upper layer $\{ -b<y<0\}$ and, for problem $L$, $\Omega_1$ is the upper layer $\{ -b<y<0\}$. Likewise, $\Omega_2$ for problem $U$ is the lower layer $\{ y<-b\}$ and, for problem $L$, $\Omega_2$ is the exterior of $\Gamma$ in the lower layer; $\lambda={\omega^2}/{g}$ is the spectral parameter. System (\ref{etiqueta_1})-(\ref{etiqueta_2_4}) should be complemented by one of the following conditions:
\begin{equation}\label{2.b.a}
\ds\frac{\partial\phi_1}{\partial n}=0\quad {\rm on}\quad\Gamma
\end{equation}
for  problem $U$, or
\begin{equation}\label{2.b.b}
\ds\frac{\partial\phi_2}{\partial n}=0\quad {\rm on}\quad\Gamma
\end{equation}
for  problem $L$.
The continuous spectrum of both  problems occupies the ray $\lambda \geq \Lambda_1$, where
\begin{equation*}
\Lambda_1 = \frac{\alpha  k  \tanh bk}{1 + \beta \tanh bk },
\quad \alpha= 1 - \beta.
\end{equation*}

The spectrum of the unperturbed problem possesses the threshold
\begin{equation*}
\Lambda _2 = k > \Lambda _1
\end{equation*}
embedded in the continuous spectrum.
The multiplicity of the spectrum for $ \Lambda _1 < \lambda < \Lambda_2$ is equal to 2 and for $\lambda > \Lambda _2$ is equal to 4.
These facts are a consequence of the following arguments: (\ref{etiqueta_1})-(\ref{etiqueta_2_4}) in the absence of the cylinder admits plane wave solutions proportional to ${\rm exp} (\pm ip x)$ with $p$ satisfying one of the dispersion relations
\begin{equation}\label{la1}
\lambda_1\left(\sqrt{k^2+p^2}\right)=\lambda
\end{equation}
or
\begin{equation}\label{la2}
\lambda_2\left(\sqrt{k^2+p^2}\right)=\lambda,
\end{equation}
where $\lambda$ is the spectral parameter entering (\ref{etiqueta_2_2}), (\ref{etiqueta_2_3}).
Here
\begin{equation}\label{la12}
\lambda_1(\tau)=\ds\frac{\alpha\tau\tanh b\tau}{1+\beta\tanh b\tau},\quad \lambda_2(\tau)=\tau.
\end{equation}
Note that $\lambda_1(k)=\Lambda_1$, $\lambda_2(k)=\Lambda_2$.\\
For further reference we indicate explicit forms of the plane waves corresponding to (\ref{la1}) and (\ref{la2}). In the first case
\begin{equation}\label{pw1}
\left\{ \begin{array}{rcl}
\phi_1 &=&g(y;\tau,\lambda)\; e^{\pm ipx}, \quad\quad\quad\quad\quad\quad -b<y<0,\\\\
\phi_2 &=& \ds\frac{1}{\tau}\;g'(-b; \tau,\lambda) \;e^{\tau(b+y)}\;e^{\pm ipx},\quad-\infty<y<-b,
\end{array}
\right.
\end{equation}
where
\begin{equation}\label{fmcg}
g(y;\tau,\lambda)=\tau\cosh\tau y+\lambda\sinh\tau y, \quad \tau=\sqrt{k^2+p^2},
\end{equation}
and $g'$ here and everywhere below  means the derivative of $g$ with respect to $y$.
In the second case
\begin{equation}\label{pw2}
\left\{ \begin{array}{rcl}
\phi_1 &=& e^{\tau y\pm ipx}, \quad -b<y<0,\\\\
\phi_2 &=& e^{\tau y\pm ipx},\quad-\infty<y<-b.
\end{array}
\right.
\end{equation}
When $\lambda<\Lambda_1$, there are no propagating modes. When $\Lambda_1<\lambda<\Lambda_2$, there are exactly two plane waves (\ref{pw1}) corresponding to a positive solution of (\ref{la1}) with respect to $p$, and when $\Lambda_2<\lambda$, then there are four plane waves (\ref{pw1}) and (\ref{pw2}) corresponding to positive solutions of (\ref{la1}) and (\ref{la2}).

Note that the function $g(y; \tau,\lambda)$ as a function of $y$ with $\tau$ and $\lambda$ fixed is not necessarily monotonic in contrast to $e^{\tau y}$; this, as we will see, is the reason for the appearance of embedded trapped modes for problem $U$ in contrast to problem $L$, for which the profile of the propagating mode in the layer containing the cylinder is monotonic.

\subsection{Main results}\label{2.2}

As is well-known, the cut-off $\Lambda_1$ of the continuous spectrum can, under a perturbation, generate a trapped mode with frequency to the left of $\Lambda_1$. We calculate the asymptotics of these trapped modes in sections\;3.4 for problem $U$ and 4.2 for problem $L$ (their existence was shown by means of numerical-analytical techniques in \cite{LintCad} and proven in \cite{NazVid2009}). These asymptotics are presented in Theorems\;\ref{1} and \ref{LN4} below; in this case we assume that
\begin{equation*}
\lambda=\Lambda_1 (1-\sigma^2),\quad \sigma\to 0\quad\text{as}\quad \varepsilon\to 0,
 \end{equation*}
  and calculate the value of $\sigma$.

 There appears a natural question whether the embedded threshold $\Lambda_2$ also generates trapped modes with
  $$\lambda=\Lambda_2(1-\sigma^2),\quad\sigma\to0\quad\text{as}\quad\ve\to0,$$
  under the perturbation. It is also well-known that, generically, it generates a resonance, which can be characterized (see, e.g., \cite{Duan}) as a pole of the reflection coefficient of the scattering problem in the complex plane, or the (complex) value of $\lambda$ for which there exist solutions of problems $U$ or $L$ satisfying the outgoing (or radiation) conditions for $|x|\to\infty$, albeit growing exponentially as $|x|\to\infty$ (plane waves with complex wavenumber). When the imaginary part of the resonance becomes equal to zero, the resonance can convert into a trapped mode. We will use the term ``resonance'' only if its imaginary part does not vanish.

As a rule, the condition of vanishing of the imaginary part of the resonance results in a  geometric condition for the perturbation (e.g., for a perturbation of the bottom of the liquid in the form of a rectangular barrier, this condition means that its width is a multiple of the wavelength of the propagating mode \cite{MIRZhJFM}). Obviously, for a thin cylinder whose width tends to $0$ as $\varepsilon\to 0$, this condition cannot be satisfied if $\varepsilon$ is sufficiently small, and hence one cylinder, seemingly, cannot generate trapped modes. This hypothesis was in part confirmed by the numerical results from \cite{LinCadScat}; the general opinion was that one thin cylinder does not trap energy, the latter always propagating to infinity along the interface. Nevertheless, it turns out that for  problem $U$ and for a very special positioning of a symmetric cylinder in the upper layer (i.e., its submergence should be in a certain correspondence with the hydrodynamic characteristics of the ``inflated'' contour $C=\{ x= X(t),\quad y=Y(t)\}$), a trapped mode can appear. For  problem $L$, the imaginary part of the resonance never vanishes due to the monotonicity of the profile of the propagating mode (\ref{pw1}) in the lower layer.

In Theorem\;\ref{2} we present the asymptotics of the resonance (its real and imaginary parts)  for problem $U$, and in Theorem\;\ref{LAN3} we present, under the condition of  vanishing of the imaginary part of the resonance, the asymptotics of the frequency of the trapped mode into which the resonance converts. In Theorem\;\ref{LN5} we present the asymptotics of the resonance for problem $L$, which, as already noted, never converts into a trapped mode. In all these cases we assume that
$
\lambda=\Lambda_2(1-\sigma^2)
$
and calculate the asymptotics of $\sigma$ as $\varepsilon\to0$.

All the theorems mentioned above are consequences of an explicit construction of the solutions of system (\ref{etiqueta_1})-(\ref{2.b.b}) which describe trapped modes (finite energy solutions) or resonances (solutions growing at infinity and satisfying the outgoing condition). The frequencies of trapped modes and resonances are defined through solutions of nonlinear equations (\ref{Asec1}), (\ref{eq.sig}), (\ref{AsecL1}), (\ref{sec eq}) for $\sigma$, which exist by the Implicit Function Theorem and are convergent power series in $\varepsilon$ and $\varepsilon_1=\varepsilon\ln\varepsilon$.\\

For applications, the leading terms of these frequencies are of principal interest and we present them in Theorems\;\ref{1}-\ref{LN5} below. 	
In order to formulate these results, we need to  introduce the following objects. Consider the exterior Neumann problem on the plane
\begin{equation}\label{NP}
\Delta \Psi = 0\quad \mathrm{in}\quad\Omega_0, \quad \left.\frac{\partial \Psi}{\partial n} \right|_C = n_2, \quad \nabla \Psi  \rightarrow 0 \quad \mathrm{as} \quad r= \sqrt{x^2 + y^2} \rightarrow \infty ,
\end{equation}
where the contour $C$ is the ``inflated'' cross-section of the cylinder, $C= \{x=X(t),  y= Y(t) \} $, $\Omega_0$ is the exterior of this contour on the plane $(x,y)$, $n_2$ is the second component of the inward-looking normal to $C$,
$n_2=  \dot{X} / \sqrt{\dot{X}^2+\dot{Y}^2}$.
Problem (\ref{NP}) describes the vertical flow of an unbounded fluid past the cylinder. The solution of this problem is unique up to an additive constant \cite{Newman} and
\begin{equation*}
\Psi = \mathrm{const} - \mu  \frac{y }{r^2}-\nu\frac{x}{r^2} + O \left( \frac{1}{r^2} \right)\quad \mathrm{as}\quad
r \to \infty.
\end{equation*}
The constants $\mu$ and $\nu$ are called \textit{strengths of the vertically and horizontally oriented dipoles} corresponding to (\ref{NP}). We indicate the formulas for $\mu$ and $\nu$ (see \cite{Newman}):
\begin{equation}\label{mu}
\mu=\frac1{2\pi}\left(S+\int\limits_{C} n_2\Psi\,dl\right),\quad \nu=\ds\frac{1}{2\pi}\ds\int\limits_{C} n_1\Psi\;dl,
\end{equation}
where $S$ is the area bounded by $C$, $dl$ is the  element of the arclength and $n_{1,2}$ are the components of the inward-looking normal to $C$. It is easy to see, integrating the form $\nabla(\Psi\nabla \Psi)$ over the exterior of $C$, that the  integral $\int_Cn_2\Psi\,dl$ in (\ref{mu}) is positive, and hence $\mu$ is also positive. For simple forms of the cylinder cross-section, the dipole strengths are well-known \cite{Newman}. For example, for the ellipse $X=a_0\cos(t+\theta_0), Y=b_0\sin(t+\theta_0)$, one has 
$$
\mu=\frac12(a_0^2\cos^2\theta_0+b_0^2\sin^2\theta_0+a_0b_0),\qquad \nu=\frac12(a_0^2-b_0^2)\sin\theta_0\cos\theta_0,
$$
and if $a_0=b_0$ (the circle)
\begin{equation}\label{circle}
\mu=a_0^2,\qquad\nu=0.\end{equation}\\

We begin with problem $U$ (consisting of (\ref{etiqueta_1})-(\ref{2.b.a})). The analysis of the exact secular equation (\ref{Asec1}) results in the following statement
\begin{teo}\label{1}
Problem $U$ possesses a finite energy solution for $\lambda=\Lambda_1(1-\sigma^2)$, where
\begin{equation}\label{sig-U}
\sigma=2\varepsilon^2 D e^{-bk}\left(S g^2\left(-a;k,\Lambda_1\right)+2\pi\mu k^{-2}\;g'^{2}\left(-a;k,\Lambda_1\right)\right)+O\left(\varepsilon^3\ln\varepsilon\right),
\end{equation}
where the positive constant $D$ is defined by (\ref{D1}).
\end{teo}
Consider now the case $\lambda=\Lambda_2(1-\sigma^2)$ (a neighborhood of the embedded threshold).
As already noted, this threshold can generate a resonance or an embedded trapped mode under a perturbation. The condition of nonvanishing of the imaginary part of the resonance guarantees its existence. We begin with the formulation of this condition. Consider the expressions
\begin{equation}\label{mathcal{R}}
\mathcal{R}=kS\; g\big(-a; \tau_1, \Lambda_2\big)+2\pi\mu\; g'\big(-a; \tau_1, \Lambda_2\big),
\end{equation}
where $\tau_1$ is the positive root of $\lambda_1(\tau)=\Lambda_2(=k)$ (it always exists and is such that $\tau_1>k$), and
\begin{equation}\label{mathcal{J}}
\mathcal{J}=2\pi\nu\sqrt{\tau_1^2-k^2}\;g\big(-a; \tau_1, \Lambda_2\big).
\end{equation}
These quantities are proportional to the leading terms of the real and imaginary parts of the ``orthogonality condition'' (see section \ref{3.6}).
We have the following
\begin{teo}\label{2}
Let $\mathcal{R}\neq0$ and/or $\mathcal{J}\neq0$. Then problem $U$ possesses a resonance for $\lambda=\Lambda_2(1-\sigma^2)$, where
\begin{align}\label{re-s}
\Re\sigma&=\ds\frac{\varepsilon^2}{2}\;D k^2 e^{-ak} \left(S+2\pi\mu\right)+O\left(\varepsilon^3\ln\varepsilon\right),
\\ \nonumber\\ \label{im-s}
\Im\sigma&=\ds\varepsilon^4\frac{\alpha k }{\beta\tau_1^3}DD_1e^{-ak-2b\tau_1} \left(\mathcal{R}^2+\mathcal{J}^2\right)+O\left(\varepsilon^5\ln\varepsilon\right),
\end{align}
where $D$ and $D_1$ are positive constants given by (\ref{RphiE}) and (\ref{FCur}), respectively.
\end{teo}

As noted above, when the cylinder is symmetric with respect to the $y$-axis (in this case, by (\ref{NP}), (\ref{mu}), $\nu=0$ and hence $\mathcal{J}=0$ by (\ref{mathcal{J}})), and the expression $\mathcal{R}$ vanishes, then, in the leading term, the imaginary part of $\sigma$ also vanishes (see (\ref{im-s})).\\
The condition $\mathcal{R}=0$ can be considered as an equation for the submergence $a$. In section \ref{3.6} we show that, for example, for sufficiently small $\alpha$, there exists a solution $a=a^{\ast}$ of the equation $\mathcal{R}=0$. In fact, the vanishing of $\mathcal{R}$ and $\mathcal{J}$ is equivalent (in the leading term) to the orthogonality condition which guarantees the exponential decrease of the corresponding solution as $|x|\to\infty$ (see section \ref{3.6}). This condition (again, in the leading term) is satisfied if $\nu=0$ and $a=a^{\ast}$. We  show that the orthogonality condition is satisfied exactly for a symmetric cylinder when the submergence $a$ is close to $a^*$,  $a=a^{\ast}\left(1+O(\varepsilon)\right)$; this value of $a$ is a solution of a certain nonlinear equation (whose leading term coincides with $\mathcal{R}=0$) which exists by the Implicit Function Theorem; moreover, under this condition $\sigma$ is purely real.

In section \ref{3.6} we prove our central result:
\begin{teo}\label{LAN3}
Let the cross-section of the cylinder be symmetric with respect to the $y$-axis and let $a^{\ast}$ be a simple zero of the function $\mathcal{R}$ considered as a function of $a$ (such $a^*$ exists if $\alpha$ is sufficiently small, see section \ref{3.6}).  Then there exists a certain value of $a=a^{\ast}\left(1+O(\varepsilon)\right)$ such that problem $U$ possesses a trapped mode for $\lambda=\Lambda_2(1-\sigma^2)$ with $\sigma$ purely real and given by the same formula (\ref{re-s}).
\end{teo}

\begin{figure}[!h]
\begin{center}
\includegraphics[width=15cm, height=5cm]{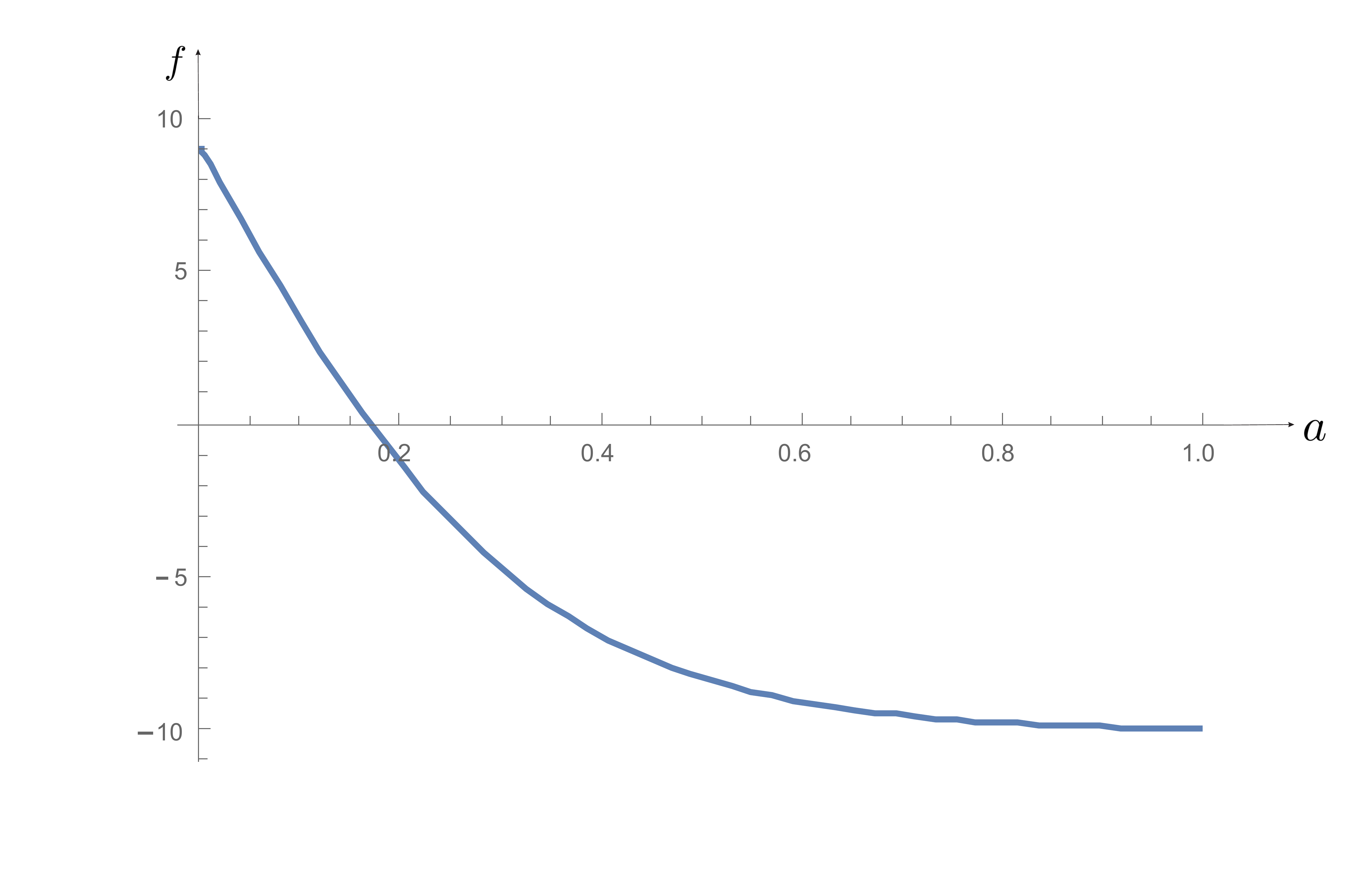}
\end{center}
\caption{Graph of $f(a),\alpha=0.5, \tau\approx3.0$}  \label{Fig3}
\end{figure}

\begin{figure}[!h]
\begin{center}
\includegraphics[width=15cm, height=5cm]{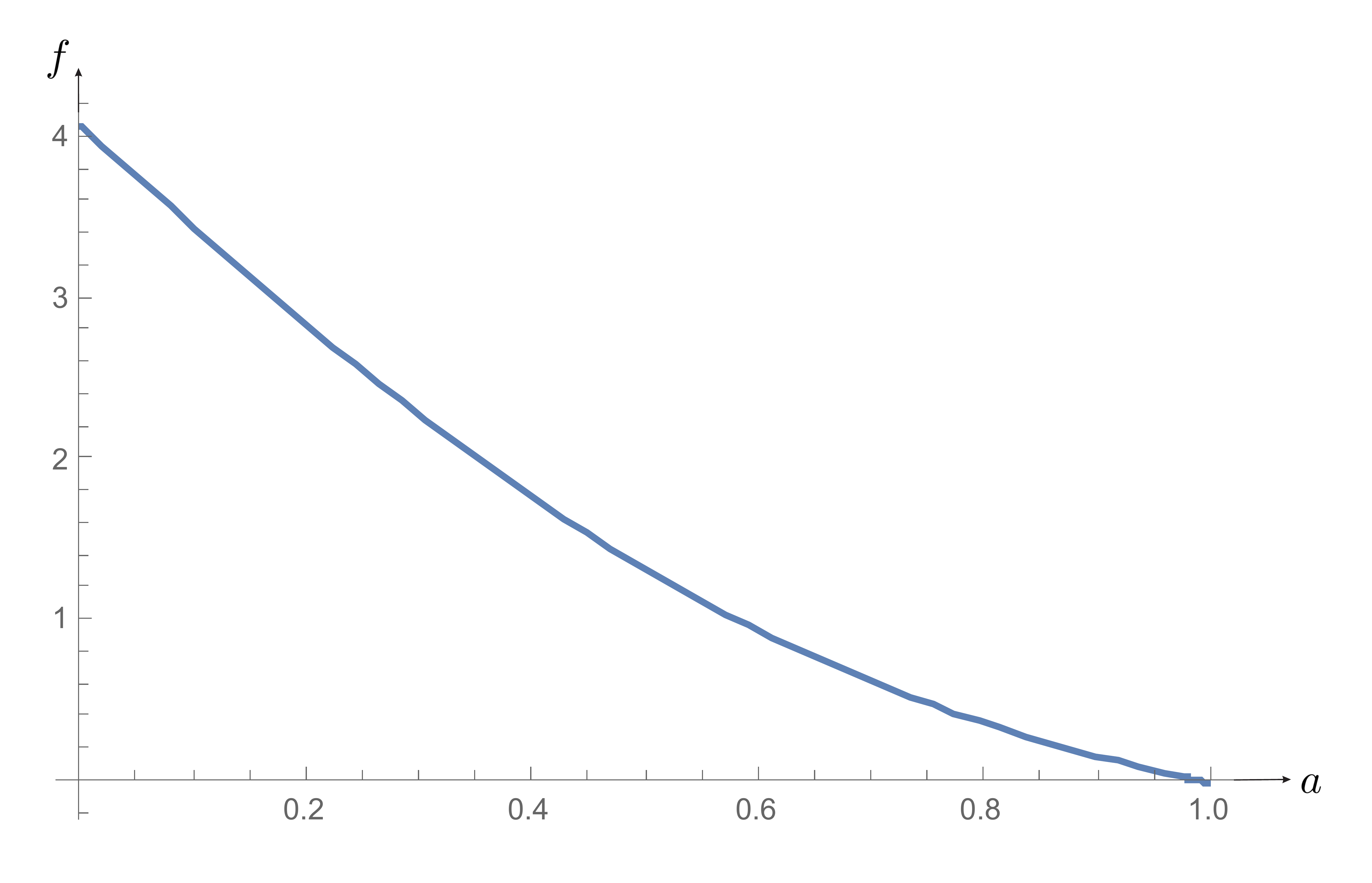}
\end{center}
\caption{Graph of $f(a),\alpha=0.91, \tau\approx1.4$}  \label{Fig4}
\end{figure}

\begin{figure}[!h]
\begin{center}
\includegraphics[width=15cm, height=5cm]{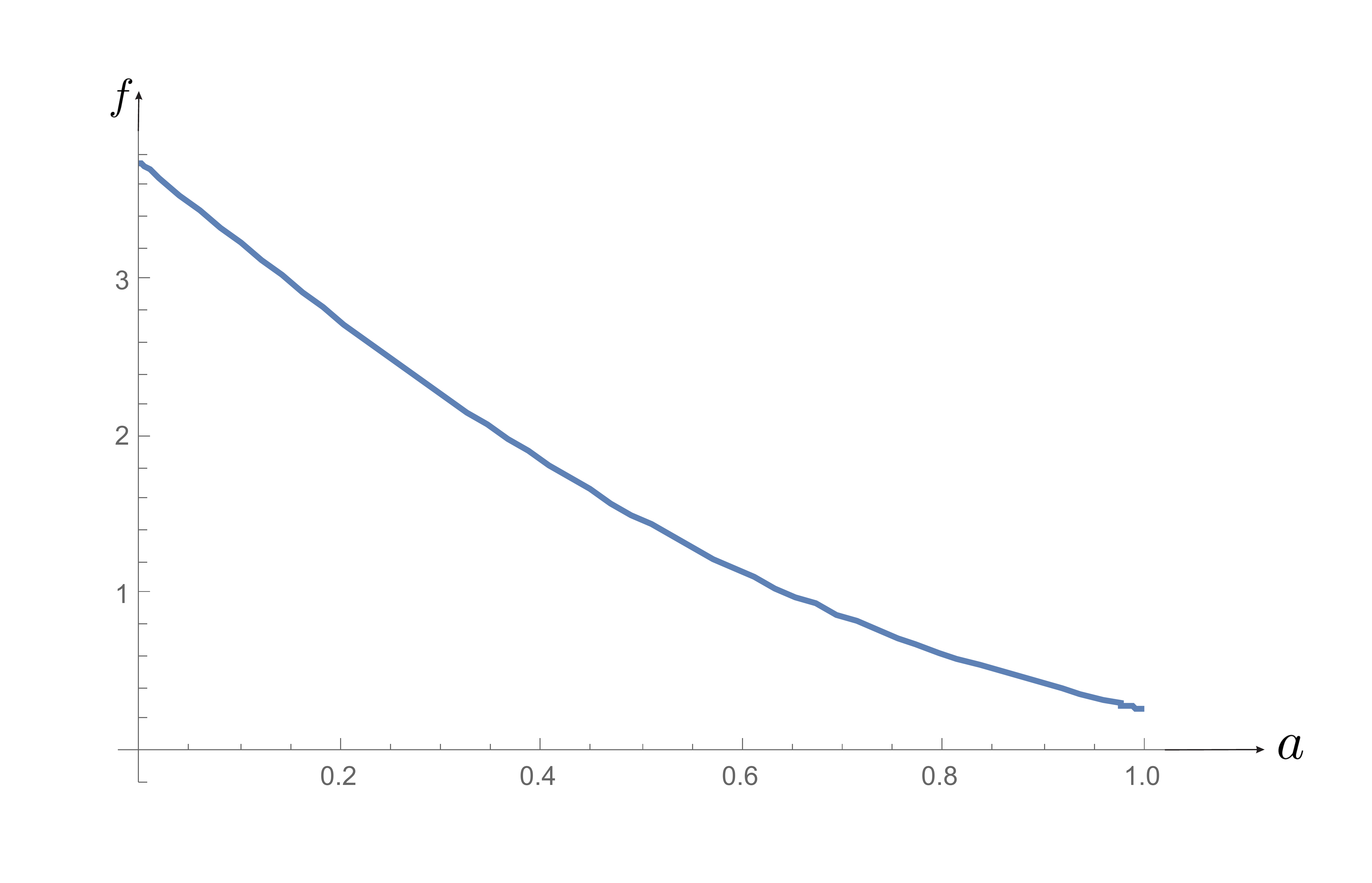}
\end{center}
\caption{Graph of $f(a),\alpha=0.97, \tau\approx1.2$}  \label{Fig5}
\end{figure}

As an illustration, we present the behavior of zeros of the function $\mathcal{R}$ as a function of the submergence $a$ for different values of the parameter $\alpha$ in the case of a circular cylinder ($X=r\cos t, Y=r\sin t, S=\pi r^2, \mu=r^2$, see (\ref{circle})). We put $k=1$ (this is equivalent to nondimensionalizing the problem, see section \ref{3.6}), $b=1, r=1$, and $\mathcal{R}$ is given by
\begin{equation}\label{num1}
\mathcal{R}=\pi \cosh a\tau(3\tau-(1+2\tau^2)\tanh a\tau),
\end{equation}
where $\tau$ is the solution of
\begin{equation}\label{taunum}1=\frac{\alpha\tau\tanh\tau}{1+(1-\alpha)\tanh\tau}.
\end{equation}
Up to a nonvanishing factor $\pi\cosh a\tau$, $\mathcal{R}$ is proportional to
\begin{equation}\label{f(a)}
f(a)=3\tau-(1+2\tau^2)\tanh a\tau,\qquad 0<a<1,
\end{equation}
and we present the graphs of this function in Fig. \ref{Fig3}, \ref{Fig4}, and \ref{Fig5} for $\alpha=0.5$, $\alpha=0.91$, and $\alpha=0.97$, respectively. The values of the corresponding  $\tau$'s (solutions of (\ref{taunum})) are approximately 3.0, 1.4, and 1.2, respectively. 

We see that, in accordance with section \ref{3.6}, the root $a^*$ of $f(a)=0$ is small for small $\alpha$ ($a^*\approx0.17$ for $\alpha=0.5$, see Fig. \ref{Fig3}) and increases as $\alpha$ increases. It almost coincides with the endpoint $a=1$ at $\alpha=0.91$ (Fig. \ref{Fig4}) and vanishes for larger values of $\alpha$ (see Fig. \ref{Fig5} for $\alpha=0.97$). For $\alpha\gtrsim0.92$ there are no zeros of $\mathcal{R}$;  hence there are no embedded trapped modes and Theorem \ref{2} holds. Thus for small $\alpha$ (almost equal densities) the trapped mode for the homogeneous case (that is, $\alpha=0$; for this value of $\alpha$ the trapped mode, which is not even embedded,  always exists) survives (although only for a special value of the submergence), and for a vanishing density of the upper layer ($\alpha\approx 1$) it disappears completely.

Let us pass now to problem $L$ (consisting of (\ref{etiqueta_1})-(\ref{etiqueta_2_4}), (\ref{2.b.b})). In this case the situation is simpler:
the first threshold $\Lambda_1$ generates a trapped mode under a perturbation and the second threshold $\Lambda_2$ generates a resonance which can never become a trapped mode since its imaginary part never vanishes.
\begin{teo}\label{LN4}
 Problem $L$ possesses a finite energy solution for $\lambda=\Lambda_1(1-\sigma^2)$, where
\begin{equation}\label{sig-L}
\sigma=\ds\frac{\varepsilon^2}{2}D e^{-ak}k\left(S+2\pi\mu\right)+O\left(\varepsilon^3\ln\varepsilon\right),
\end{equation}
and the positive constant D is defined by (\ref{Dq}).
\end{teo}
\begin{teo}\label{LN5}
Problem $L$ possesses a resonance for $\lambda=\Lambda_2\left(1-\sigma^2\right)$, where
\begin{align}\label{re-s-L}
\Re\sigma&=\ds\frac{\varepsilon^2}{2}D e^{-ak}k\left(S+2\pi\mu\right)+O\left(\varepsilon^3\ln\varepsilon\right),\\\label{im-s-L}
\Im\sigma&=\ds\frac{\varepsilon^4}{4}\ds\frac{k}{\tau_1}DD_1 e^{-2a\tau_1-ak}\left(\left(kS+2\pi \tau_1\mu\right)^2+\left(2\pi \nu\right)^2({\tau_1^2-k^2})\right)+O\left(\varepsilon^5\ln\varepsilon\right)\neq0,
\end{align}
and the positive constants $D$ and $D_1$ are defined by (\ref{Dq1}) and (\ref{gammaL}), respectively.
\end{teo}
\begin{obs}\label{RT}
Note that imaginary parts of $\sigma$ in Theorems \ref{2} and \ref{LN5} govern the temporal behavior of resonances for large values of time $t$. We have
$$\exp\{i(kz-\omega t)\}\phi_{1,2}\sim\exp\left\{-(kg)^{1/2}(\Re\sigma)(\Im\sigma)\, t\right\}
$$
and the decay rate is very low ($\sim\ve^6$).
\end{obs}
\begin{obs}\label{difsym}
In Theorems \ref{1}-\ref{LN5} identical symbols denote different (but similar) objects. This is explicitly indicated in the corresponding references. In what follows, we will use the same convention.
\end{obs}
\subsection{Limiting cases}\label{2.3}
As already mentioned, the limiting cases $\beta\to1$ (almost equal densities), $\beta\to 0$ (the upper layer disappears), and even $\varepsilon\to0$ (the cylinder disappears) are all singular; therefore, our formulas for discrete eigenvalues do not need to pass into those of, e.g., \cite{NazVid} as $\beta\to0,1$. Nevertheless, we indicate the limiting behavior of our formulas under some of  these passages to the limit in the cases when the corresponding limits can be compared with known results. It turns out that qualitatively they are similar to those of \cite{NazVid}, while the asymptotics of the resonances and embedded trapped modes (not considered in \cite{NazVid}) do pass into the eigenvalues for a homogeneous fluid.
\begin{enumerate}
\item
 Formula (\ref{sig-U}) from Theorem \ref{1} implies the following limiting behavior of the discrete eigenvalue:\\
(a) if $\alpha\to0$, then the threshold $\Lambda _1\sim\alpha$ and $\sigma\sim\varepsilon^2$; hence the eigenvalue lies to the left of $\Lambda_1$ at a distance $\sim\alpha\varepsilon^4$.\\
(b) if $\beta\to0$, then $\Lambda_1=k\tanh kb+O(\beta)$ and $\sigma\sim\varepsilon^2$; hence the eigenvalue lies to the left of $\Lambda_1$ at a distance $\sim\varepsilon^4$.
\item
 The same qualitative behavior is observed for the discrete eigenvalue described by Theorem \ref{LN4} for problem $L$ (see (\ref{sig-L})).
 \item
 Formulas (\ref{re-s}) and (\ref{im-s}) from Theorem \ref{2} for the resonance imply that
 if $\alpha\to0$, then $\Im\sigma\sim\alpha$ and
\begin{equation}\label{Ur}
\Re \sigma=\frac{\varepsilon^2}{\sqrt2}e^{-2ak}k^2(S+2\pi\mu)+O(\alpha\varepsilon^2)+ O(\ve^3\ln\ve);
\end{equation}
the last formula coincides with the results from \cite{Ursell} (for a circle), \cite{McIver} (for a symmetric body) and \cite{GarZhev} (for a nonsymmetric body).
 \item Formulas (\ref{re-s-L}) and (\ref{im-s-L}) from Theorem \ref{LN5} have the same qualitative behavior as in item 3. In formula (\ref{Ur}) the exponential $e^{-2ak}$ changes to $e^{-2(a+b)k}$ (the total submergence of the cylinder is equal to $a+b$).
     \end{enumerate}
\section{Cylinder in the upper layer}
\setcounter{equation}{0}

\subsection{System of integral equations}\label{3.1}
Throughout the text an integral without limits means that the integration is carried out over the whole real axis. We use the following Fourier transform formula:
\begin{equation}\label{Fourier}
\tilde f(p)=F_{x\to p}[f(x)]=\int e^{-ipx}f(x)\,dx;
\end{equation}
the inverse Fourier transform is given by
$$f(x)=F_{p\to x}^{-1}[\tilde f(p)]=\frac1{2\pi}\int e^{ipx}\tilde f(p)\,dp.$$
Introduce the functions $\varphi$, $\psi$ and $\theta$ by the formulas
\begin{equation}\label{Def}
\vf:=\left.\phi_{1}\right\vert_{y=0},\qquad \psi:=\left.\phi_{1y}\right\vert_{y=-b}=\left.\phi_{2y}\right\vert_{y=-b},\qquad \theta:=\left.\phi_1\right\vert_{\Gamma}.
\end{equation}
If these functions are known, then $\phi_2$ can be reconstructed as the solution of the problem consisting of (\ref{etiqueta_1_2}) and the condition $\left.\phi_{2y}\right\vert_{y=-b}=\psi$ by means of the Fourier transform. Indeed, denoting the Fourier transform of $\phi_2$ with respect to $x$ by $\tilde\phi_2(p,y)$, we have
$$
\tilde{\phi}_2=\ds\frac{1}{\tau} e^{(b+y)\tau}\;\tilde{\psi},\qquad \tau=\sqrt{k^2+p^2},
$$
and hence (by (\ref{etiqueta_2_3}) and (\ref{etiqueta_2_4}))
\begin{equation}\label{phi1 psi}
\left.\tilde{\phi}_1\right\vert_{y=-b}=\ds\frac{1}{\beta}\Big(\ds\frac{1}{\tau}-\ds\frac{\alpha}{\lambda}\Big)\tilde{\psi}.
\end{equation}
Therefore, $\phi_1$ and its normal derivatives on the boundary of $\Omega_1$ are given by (\ref{phi1 psi}), $\left.\phi_{1y}\right\vert_{y=-b}=\psi$  (by (\ref{etiqueta_2_4})), $\left.\frac{\partial \phi_1}{\partial n}\right\vert_{\Gamma}=0$ (by (\ref{2.b.a})), $\left.\phi_1\right\vert_{\Gamma}=\theta$ (by definition (\ref{Def})), $\left.\phi_1\right\vert_{\Gamma_F}=\varphi$ (by definition (\ref{Def})), $\left.\phi_{1y}\right\vert_{\Gamma_F}=\lambda \varphi$ (by (\ref{etiqueta_2_2})).
Hence, by means of the Green formula (\ref{GF}), $\phi_1$ can be reconstructed in the whole upper layer.

As shown in Appendix\;1, system (\ref{etiqueta_1})-(\ref{2.b.a}) can be reduced to three integral equations for $\varphi, \psi, \theta$. They have the form
\begin{equation}\label{M4}
\displaystyle\frac{(\tau-\lambda)}{\tau}\tilde{\varphi}+\displaystyle\frac{(\tau-\lambda)}
{\tau}\;\displaystyle\frac{\alpha}{\beta\lambda}\;e^{-b\tau}\;\tilde{\psi}=\varepsilon\displaystyle\int\limits_{-\pi}^{\pi} M_4(p,t)\;\theta(t)\; dt,
\end{equation}
\begin{equation}
 \label{M5}
-\displaystyle\frac{(\tau+\lambda)}{\tau}\;e^{-b\tau}\;\tilde{\varphi}+\displaystyle\frac{\tau+\lambda}{\tau}\;
\displaystyle\frac{1}{\lambda}\left(1-\displaystyle\frac{1}{\beta}\;\displaystyle\frac{\tau-\lambda}{\tau+\lambda}\right)
\;\tilde{\psi}=\varepsilon\displaystyle\int\limits_{-\pi}^{\pi} M_5(p,t)\;\theta(t)\; dt
\end{equation}
\begin{equation} \label{M1}
\theta+\hat{M}_1\theta=\displaystyle\int M_2(t,p)\;\tilde{\varphi}(p)\, dp+\displaystyle\int M_3(t,p)\;\tilde{\psi}(p)\, dp,
\end{equation}
the operator $\hat{M}_1$ is given by the formula
\begin{equation}\label{M_1the}
\hat{M}_1\theta(t)=\ds\int\limits_{-\pi}^{\pi} M_1(t,s)\;\theta(s)\;ds,
\end{equation}
where
\begin{equation}\label{M_1'}
M_1(t,s)=\ds\frac{\varepsilon k}{\pi} K'_0\left(\varepsilon k \left|{\bf r}(s)-{\bf r}(t)\right|\right)\;\ds\frac{\left({\bf r}(s)-{\bf r}(t)\right)\cdot{\bf m}(s)}{\left|{\bf r}(s)-{\bf r}(t)\right|},
\end{equation}
\begin{equation}\label{bfmr}
{\bf r}(s)=(X(s),Y(s)),\qquad {\bf m}(s)=(-\dot{Y}(s),\dot{X}(s)).
\end{equation}
The other operators appearing in (\ref{M4})-(\ref{M1}) are given by
\begin{equation}\label{M2}
M_2(t,p)=\displaystyle\frac{1}{2\pi}\;\displaystyle\frac{\tau+\lambda}{\tau}\;e^{-(a-\varepsilon Y)\tau+ip\varepsilon X},\quad
M_3(t,p)=-\displaystyle\frac{1}{2\pi}\;\displaystyle\frac{\alpha}{\beta\lambda}\;\displaystyle\frac{\tau-\lambda}
{\tau}
e^{-(c+\varepsilon Y)\tau+ip\varepsilon X}
\end{equation}
\begin{equation}\label{M_4 1}
M_4(p,t)=\left(-\displaystyle\frac{ip}{\tau}\dot{Y}-\dot{X}\right) e^{-(a-\varepsilon Y)\tau-ip\varepsilon X},\quad
M_5(p,t)=\left(-\displaystyle\frac{ip}{\tau}\dot{Y}+\dot{X}\right) e^{-(c+\varepsilon Y)\tau-ip\varepsilon X},
\end{equation}
where $c=b-a$.
Note that all the kernels depend also on $\varepsilon$; in what follows we will usually omit $\varepsilon$ as  an argument of the corresponding functions.

\subsection{Reduction to two equations}\label{3.2}

Let us analyze the structure of the integral operator $M_1$.
We have
$$
K_0'(r)=-\ds\frac{1}{r}+\ds\frac{1}{2}(1-\gamma) r-\ds\frac{1}{2} r\ln\ds\frac{r}{2}+O(r^3\ln r).
$$
Hence,
\begin{equation}\label{M_1}
M_1(t,s)=M_1^{(0)}(t,s)+\varepsilon^2 M_1^{(1)}(t,s,\varepsilon,\varepsilon\ln\varepsilon)+\varepsilon^2 \ln\varepsilon\; M_1^{(2)} (t,s,\varepsilon,\varepsilon\ln\varepsilon),
\end{equation}
where
\begin{equation}\label{M_1^{0}}
M_1^{(0)}(t,s)=-2\ds\frac{\partial G_0\left({\bf r}(s)-{\bf r}(t)\right)}{\partial n}\left|{\bf m}(s)\right|
\end{equation}
with
$$
G_0(x,y)=\ds\frac{1}{2\pi}\ln r, \quad r=\sqrt{x^2+y^2},
$$
and $\ds\frac{\partial}{\partial n}$ is the derivative along the inward-looking normal to $C$, the normal derivative is calculated at the point $\left(X(s), Y(s)\right)$. The kernels $M_1^{(1)}$ and $M_1^{(2)}$ are smooth in $t,s$ and analytic in $\varepsilon, \varepsilon_1=\varepsilon\ln\varepsilon$.
It is well-known that the operator $1+\hat{M}_1^{(0)}$ is invertible  \cite{Petrovsky} (e.g., in the space $C[-\pi,\pi]$ of continuous functions with the sup-norm), and hence $1+\hat{M}_1$ is also invertible by (\ref{M_1}). We will denote
\begin{equation}\label{NN0}
\hat N=\left(1+\hat M_1\right)^{-1},\qquad \hat N_0=\left(1+\hat M_1^{(0)}\right)^{-1}.
\end{equation}
Thus (\ref{M4})-(\ref{M1}) reduces to the following system for $\tilde{\varphi}, \tilde{\psi}$:
\begin{align}\label{ml}
B\begin{pmatrix}
\tilde{\varphi}\\
\tilde{\psi}
\end{pmatrix}=
\varepsilon\begin{pmatrix}
\hat{M}_4\\
\hat{M}_5
\end{pmatrix}\hat N\left(\hat{M}_2, \hat{M}_3\right)\begin{pmatrix}
\tilde{\varphi}\\
\tilde{\psi}
\end{pmatrix},
\end{align}
where
\begin{equation}\label{B}
B(\tau)=\begin{pmatrix}
\displaystyle\frac{\tau-\lambda}{\tau}&&\displaystyle\frac{\tau-\lambda}{\tau}\;\displaystyle\frac{\alpha}{\beta\lambda}
\;e^{-b\tau}\\\\
-\displaystyle\frac{\tau+\lambda}{\tau}\;e^{-b\tau}&&\displaystyle\frac{\tau+\lambda}{\tau}
\;\displaystyle\frac{1}{\lambda}\left(1-\displaystyle\frac{1}{\beta}\;\displaystyle\frac{\tau-\lambda}{\tau+\lambda}\right)
\end{pmatrix},\qquad \tau=\sqrt{k^2+p^2}.
\end{equation}
We use here the following notation:
$$ \left(\hat{M}_2, \hat{M}_3\right)\begin{pmatrix}
\tilde{\varphi}\\
\tilde{\psi}
\end{pmatrix}=\hat M_2\tilde \varphi+\hat M_3\tilde\psi,\qquad \begin{pmatrix}
\hat{M}_4\\
\hat{M}_5
\end{pmatrix}\theta=\begin{pmatrix}
\hat{M}_4\theta\\
\hat{M}_5\theta
\end{pmatrix}.
$$

It is convenient to introduce formally a new unknown ${\bf A}(p)$ by the formula
\begin{equation}\label{phi psi}
\begin{pmatrix}
\tilde{\varphi}\\
\tilde{\psi}
\end{pmatrix}=B^{-1}{\bf A},\qquad {\bf A}=\begin{pmatrix}
A_1(p)\\
A_2(p)
\end{pmatrix}.
\end{equation}
All the solutions constructed in what follows will have the property that the components of the corresponding ${\bf A}$'s
belong to  the Banach space of bounded analytic functions in a sufficiently narrow strip $\mathcal{S}$ containing the real axis with the sup-norm; we will denote it by $\mathscr{A}(\mathcal{S})$. The width of $\mathcal{S}$ is chosen so that it does not contain zeros of $\det B$ which are bounded away from the real axis and does contain zeros of $\det B$ which are close to it and are described in the next section. Moreover, the width of $\mathcal{S}$  does not depend on $\ve$.

We have
\begin{equation}\label{B-1}
B^{-1}=\ds\frac{1}{\det B}\operatorname{adj}B,
\end{equation}
where
\begin{equation*}
\operatorname{adj}B=\begin{pmatrix}
\displaystyle\frac{\tau+\lambda}{\tau}\;\displaystyle\frac{1}{\lambda}\left(1-\displaystyle\frac{1}{\beta}\;\displaystyle
\frac{\tau-\lambda}{\tau+\lambda}\right)&&-\displaystyle\frac{\tau-\lambda}{\tau}\;\displaystyle
\frac{\alpha}{\beta\lambda}\;e^{-b\tau}\\\\
\displaystyle\frac{\tau+\lambda}{\tau}\;e^{-b\tau}&& \displaystyle\frac{\tau-\lambda}{\tau}
\end{pmatrix},
\end{equation*}
\begin{equation*}
\det B =\ds\frac{\tau^2-\lambda^2}{\lambda\tau^2}\left(1-\displaystyle\frac{1}{\beta}\;\displaystyle
\frac{\tau-\lambda}{\tau+\lambda}+\displaystyle \displaystyle\frac{\alpha}{\beta}e^{-2b\tau}\right).
\end{equation*}
Here $\operatorname{adj} B$ means the adjugate matrix.
After some algebraic manipulations, we obtain
\begin{equation*}
\det B=-\ds\frac{Q(\tau)}{\lambda}\left(\lambda_1(\tau)-\lambda\right) \left(\lambda_2(\tau)-\lambda\right),
\end{equation*}
where $Q(\tau)$ is given by
\begin{equation}\label{QU}
Q(\tau)=\ds\frac{2}{\beta\tau^2}\;e^{-b\tau}\left(\cosh b\tau+\beta\sinh b\tau\right)
\end{equation}
and $\lambda_{1,2}(\tau)$ are given by (\ref{la12}).

\subsection{Zeros of $\det B$}\label{3.3}

 Since zeros of $\det B$ coincide with the poles of $\tilde\vf$, $\tilde\psi$ by (\ref{phi psi}), we will have to localize these zeros   in the strip $\mathcal{S}$  for the values of $\lambda$ close to the thresholds $\Lambda_{1,2}$. Consider first the case $\lambda=\Lambda_1(1-\sigma^2)$, $\sigma\to 0$ as $\ve\to0$. Obviously, $Q(\tau(p))$ does not vanish in a sufficiently narrow strip $\mathcal{S}$, and the same is true for $\lambda_2(\tau(p))-\lambda$. Hence $\det B=0$ in $\mathcal{S}$ only if $\lambda_1(\tau(p))=\lambda$. The last equation is satisfied at the points
\begin{equation}\label{pOG}
p=\pm ip_{01}(\sigma),\quad p_{01}(\sigma)= q_1\sigma+O(\sigma^2),
\quad
q_1 =\sqrt{\ds{2k\Lambda_1}/{\lambda'_1(k)}}.
\end{equation}
All the other zeros of $\lambda_1(\tau(p))-\lambda$ lie outside $\mathcal{S}$; the same is true, by the above argument, for the zeros of $\det B$.\\

Consider now the case $\lambda=\Lambda_2(1-\sigma^2)$, $\sigma\to0$ as $\varepsilon\to 0$, $\Re\sigma>0$.
As we already know, $\det B(\tau)$ vanishes only at the points $\tau$ such that
\begin{equation}\label{lam1}
\lambda_1(\tau)=\lambda
\end{equation}
or
\begin{equation}\label{lam2}
\lambda_2(\tau)=\lambda.
\end{equation}
 For any real $\lambda$ satisfying $\lambda>\Lambda_1$,   equation (\ref{lam1}) has exactly one simple positive root which is greater than  $k$.

Assume now that $\sigma$ is small and (possibly) complex with  positive real and imaginary parts (we will see later that this is the case), and consider  equation (\ref{lam1}). This equation will also possess one solution $\tau=\tau_1^*$ which coincides with $\tau_1$ (introduced in (\ref{mathcal{R}})) for $\sigma=0$.\\
Let us calculate the asymptotics of this solution for small (possibly complex) $\sigma$. We have for $\tau_1^*$:
\begin{equation}\label{tansig}
\tau_1^*=\tau_1+\tau_1^1\sigma^2+\tau_1^2\sigma^4+\cdots,
\end{equation}
where $\tau_1$ satisfies $\Lambda_2=\lambda_1(\tau_1)$, and $\tau_1^1, \tau_1^2,\dots$ should be determined. Substituting in (\ref{lam1}), we obtain
$$
\tau_1^1=-\ds\frac{k}{\lambda'_1(\tau_1)}<0.
$$
Similarly, for the corresponding wavenumber $p_1=\sqrt{(\tau_1^*)^2-k^2}$ we have
\begin{equation}\label{p^1}
p_1=\sqrt{(\tau_1)^2-k^2}\left(1+\ds\frac{\tau_1 \tau_1^1}{\left((\tau_1)^2-k^2\right)^{{3}/{2}}}\;\sigma^2+\cdots\right)
\end{equation}
and hence $\Im p_1<0$. Thus $\det B$ vanishes at the points $\pm p_1$ (see Fig.\;\ref{X1}); $p_1(\sigma)=p_1^0+O(\sigma^2)$,  $p_1^0=\sqrt{(\tau_1)^2-k^2}$.\\

Consider now equation (\ref{lam2}). It can be rewritten as
$
p^2=\lambda^2-k^2
$
and for $\lambda=\Lambda_2(1-\sigma^2)=k(1-\sigma^2)$ has the form
$
p^2=-2k^2\sigma^2+k^2\sigma^4.
$
The last equation has, for small $\sigma$, the two roots
\begin{equation}\label{p phiphi}
p=\pm ip_{02}(\sigma), \quad p_{02}=k\sigma \sqrt{2-\sigma^2}=q_{2}\sigma+O(\sigma^3),\quad q_{2}=k\sqrt{2}.
\end{equation}
Thus $\det B$ has also two simple zeros at $p=\pm ip_{02}$. All the other zeros of $\det B$ lie outside $\mathcal{S}$.


\subsection{Discrete eigenvalue}\label{3.4}
\begin{figure}[!h]
\begin{center}
\includegraphics[width=15cm, height=2.33cm]{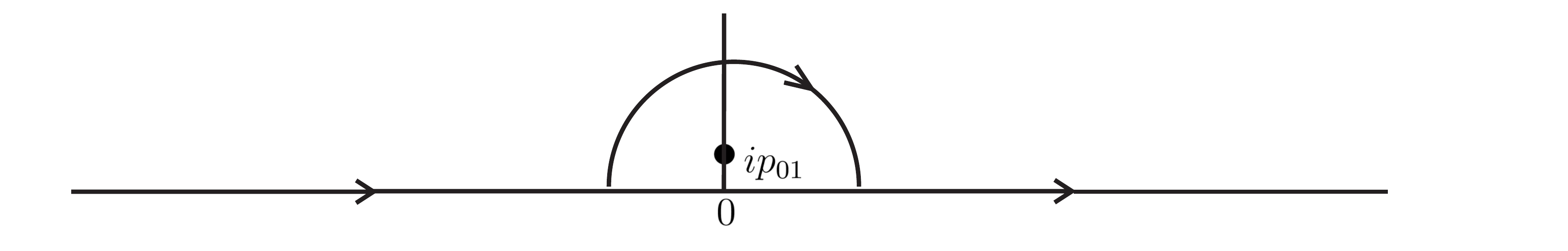}
\end{center}
\caption{Contour $\gamma'$}  \label{ConGama}
\end{figure}

In this section we present the proof of Theorem \ref{1}.
First, we consider the case of the discrete eigenvalue which appears under the perturbation to the left of the first cut-off $\Lambda_1$. Thus we look for the eigenvalue $\lambda$ in the form
\begin{equation}\label{laa1}
\lambda=\Lambda_1(1-\sigma^2),\quad\sigma\to0\quad\text{as}\quad\varepsilon\to0.
\end{equation}
We will construct a solution of (\ref{ml}) with $\lambda$ given by (\ref{laa1}), that is, we will construct the pair $(\sigma, {\bf A})$ which satisfies (\ref{ml}), (\ref{phi psi}).\\
Here the components of ${\bf A}$ belong to $\mathscr{A}(\mathcal{S})$, the norm of ${\bf A}$
being the maximum of the norms of the components.
We look for the solution of (\ref{ml}) in the form (\ref{phi psi})
with ${\bf A}=(A_1(p), A_2(p))^{\top}$ a new unknown.
Substituting (\ref{phi psi}) in (\ref{ml}), we obtain
\begin{equation}\label{eqA1}
{\bf A}=\varepsilon \hat{M}\hat{K}{\bf A},
\end{equation}
where
\begin{equation}\label{hatM1}
\hat{M}=\begin{pmatrix}
\hat{M}_4\\
\hat{M}_5
\end{pmatrix}\left(1+\hat{M}_1\right)^{-1},\quad \hat{K}{\bf A}:=\left(\hat{M}_2,\hat{M}_3\right) B^{-1}{\bf A}.
\end{equation}
In the case under consideration $\det B$ has two simple zeros in $\mathcal{S}$ given by (\ref{pOG}) and hence the operator $\hat{K}$ in the right-hand side of (\ref{eqA1}) is not bounded uniformly in $\sigma$ since  the points $\pm ip_{01}$ are close to the real axis.
Hence, the operator in the right-hand side of (\ref{eqA1}) is not bounded uniformly with respect to $\sigma$. We remedy this situation by subtracting and adding the principal part (with respect to $\sigma$) of the residue of the integrand of $\left(\hat{M}_2,\hat{M}_3\right) B^{-1}{\bf A}$. More precisely, the following lemma holds. Introduce the following operator from $\mathscr{A}^2=\mathscr{A}\times\mathscr{A}$ to $\mathscr{A}$:
\begin{equation}
 \label{Ort}
\Ort({\bf A})=\ds\frac{\tau+\lambda}{\tau}\;e^{-b\tau} A_1(p)+\ds\frac{\tau-\lambda}{\tau}\;A_2(p).
\end{equation}
\begin{lem}\label{Op Reg}
Let $\lambda=\Lambda_1(1-\sigma^2)$ and let the operator $\hat{K}$ acting from the space $\mathscr{A}^2$ to the space  $C[-\pi,\pi]$ be given by (\ref{hatM1}),
\begin{equation}\label{KA}
\hat{K} {\bf A}=\ds\int\left(M_2(t,p), M_3(t,p)\right) B^{-1} {\bf A}\,dp.
\end{equation}
Then the operator
\begin{equation}\label{T0A}
\hat{T}_0{\bf A}:=\hat{K}{\bf A}-\ds\frac{1}{\sigma}\left(\left.\operatorname{Ort}({\bf A})\right\vert_{p=0}\right)R_0(t),
\end{equation}
where
\begin{equation}\label{Rphi}
R_0(t)=D\left((k+\Lambda_1)e^{-ak+\varepsilon k Y}+(k-\Lambda_1)e^{ak-\varepsilon k Y}\right),
\end{equation}
\begin{align}\label{D1}
D&=\ds\frac{\alpha}{\beta}\;e^{-bk}\;\ds\frac{1}{Q\left(k\right)\;\lambda'_1(k)\left(\Lambda_2-\Lambda_1\right) q_1}>0,
\end{align}
is bounded uniformly in $\sigma$ for small $\sigma$. Moreover, if ${\bf A}$ is analytic in $\varepsilon, \sigma$, then $\hat{T}_0{\bf A}$ also is. Here $Q(\tau)$ is defined by (\ref{QU}) and $q_1$ by (\ref{pOG}).
\end{lem}
{\bf Proof}. The proof consists in changing the contour of integration in (\ref{KA}) to the one shown in Fig. \ref{ConGama} and the application of the Cauchy residue theorem; the second summand in the right-hand side of (\ref{T0A}) is simply the principal part of the Laurent series with respect to $\sigma$ of $2\pi i$ times the residue at $p=ip_0$. The analyticity follows from the fact  that the integral over $\gamma'$ (which coincides with $\hat{T}_0{\bf A}$ up to an analytic function) is analytic.$\qquad\blacksquare$

\medskip
By (\ref{T0A}) equation (\ref{eqA1}) takes de form
\begin{equation*}
{\bf A}=\varepsilon \hat{M}\hat{T}_0 {\bf A}+\ds\frac{\varepsilon}{\sigma}\left(\left.\operatorname{Ort}({\bf A})\right\vert_{p=0}\right)\hat{M} R_0,
\end{equation*}
and, by Lemma\;\ref{Op Reg}, the operator $\hat{T}:=\hat{M}\hat{T}_0$ is bounded. Hence
\begin{equation}\label{Asec3}
{\bf A}=\ds\frac{\varepsilon}{\sigma}\left(\left.\operatorname{Ort}({\bf A})\right\vert_{p=0}\right)\left(1-\varepsilon\hat{T}\right)^{-1}\hat{M} R_0.
\end{equation}
We will prove the existence of the eigenvalue $\lambda=\Lambda_1(1-\sigma^2)$. To this end we will find the value of $\sigma$ and the ``eigenfunction'' ${\bf A}$.

Applying the operator Ort given by (\ref{Ort}) to  equation (\ref{Asec3}) and calculating the value of the result at $p=0$, we see that the constant ${\rm Ort} ({\bf A})\Big\vert_{p=0}$ cancels out and, multiplying by $\sigma$, we obtain a secular equation for $\sigma$:
\begin{equation}\label{Asec1}
\sigma=\varepsilon F(\varepsilon,\varepsilon_1,\sigma),
\end{equation}
where
\begin{equation}\label{EP}
F(\varepsilon,\varepsilon_1,\sigma)=\left.{\Ort}\left(\left(1-\varepsilon\hat{T}\right)^{-1}\hat{M} R_0\right)\right\vert_{p=0}.
\end{equation}
Obviously, $F(\varepsilon,\varepsilon_1,\sigma)$ is analytic in $\varepsilon$, $\ve_1$ and $\sigma$, and hence (\ref{Asec1}) possesses, by the Implicit Function Theorem, an analytic in $\varepsilon, \varepsilon_1$, solution
\begin{equation}\label{sig varep}
\sigma=\sigma(\varepsilon, \varepsilon_1).
\end{equation}
The fact that ${\bf A}$ given by (\ref{Asec3}) belongs to $\mathscr{A}^2$ follows immediately from the explicit formulas for the operators $\hat{M}_{j}$; moreover, the same formulas guarantee that ${\bf A}$ decreases exponentially as $|p|\to\infty$.
Thus, taking into account the fact that ${\bf A}$ defines, through $\tilde{\varphi}, \tilde{\psi}, \theta$, a finite energy solution of problem $U$, we have proven the first statement of Theorem\;\ref{1}. In order to conclude the proof of this theorem, we will calculate the asymptotics of the solution of (\ref{Asec1}).
Let us calculate the leading term in (\ref{sig varep}). We will see that $\sigma=O(\varepsilon^2)$. This fact is the consequence of the following
\begin{lem}\label{hat{M}R_0 O}
We have
$
\hat{M} R_0=O(\varepsilon).
$
\end{lem}
{\bf Proof}. By (\ref{Rphi}),
$
R_0(t)={2Dg}+O(\varepsilon);$
we omit the arguments $\left(-a;k,\Lambda_1\right)$ of $g$ and $g'$ up to the end of this subsection.
By (\ref{M_1}), (\ref{NN0})
\begin{equation}\label{NNphi}
\hat{N}=\hat{N}_0+O\left(\varepsilon^2\ln\varepsilon\right).
\end{equation}
By (\ref{M_4 1}),
$$
M_4=
\left(-\ds\frac{ip}{\tau}\dot{Y}-\dot{X}\right)\left(e^{-a\tau}+O(\varepsilon)\right),\quad M_5=
\left(-\ds\frac{ip}{\tau}\dot{Y}+\dot{X}\right)\left(e^{-c\tau}+O(\varepsilon)\right),
$$
and we see, using (\ref{lemma45}), that in the leading term the expression $\hat{M} R_0$ vanishes since
$
\int_{-\pi}^{\pi}\dot{Y}\,dt=\int_{-\pi}^{\pi}\dot{X}\,dt=0.~~~~~~\blacksquare
$

\medskip
This lemma shows that, in order to calculate the leading term of $\sigma(\varepsilon, \varepsilon_1)$, we have to consider only the expression $\left.{\Ort} \left(\hat{M} R_0\right)\right\vert_{p=0}$ in (\ref{EP}).\\
Indeed, we have
\begin{equation}\label{TrasF}
F=\left.{\Ort} \left(\hat{M} R_0\right)\right\vert_{p=0}+\varepsilon\left. {\Ort} \left(\hat{T} \hat{M} R_0\right)\right\vert_{p=0}+O(\varepsilon^2),
\end{equation}
and, by Lemma\;\ref{hat{M}R_0 O}, the second term in (\ref{TrasF}) is $O(\varepsilon^2)$.

We have for $\hat{M} R_0$
\begin{equation}\label{MR phiphi}
\left.\hat{M} R_0\right\vert_{p=0}= \int\limits_{-\pi}^{\pi}\begin{pmatrix}
-\dot{X} e^{-(a-\varepsilon Y)k}\\
\dot{X} e^{-(c+\varepsilon Y)k}
\end{pmatrix} \hat{N} R_0\;dt.
\end{equation}
Using the expansions
$e^{\pm\varepsilon kY}=1\pm \varepsilon kY+O(\varepsilon^2),
$
formulas
 (\ref{Rphi}), (\ref{NNphi}),
 (\ref{lemma45}) and Proposition\;\ref{NN}, we obtain
\begin{equation}\label{MR' phiphi}
\left.\hat{M} R_0\right\vert_{p=0}=\varepsilon D\begin{pmatrix}
e^{-ak}\left(k S g+2\pi\mu g'\right)\\
e^{-ck}\left(k S g-2\pi\mu g'\right)
\end{pmatrix}+O\left(\varepsilon^2\ln\varepsilon+\sigma^2\right).
\end{equation}
Substituting (\ref{MR' phiphi}) in (\ref{TrasF}),  we obtain, by the definition of $g$ (\ref{fmcg}) and (\ref{TrasF}),
\begin{align*}
F(\varepsilon, \varepsilon_1,\sigma)&=
\ds\frac{\varepsilon D e^{-kb}}{k}\Big(\left(k+\Lambda_1\right) e^{-ak}\left(k Sg+2\pi\mu g'\right)\\
&\qquad\qquad\qquad+ \left(k-\Lambda_1\right) e^{ak}\left(k Sg-2\pi\mu g'\right)\Big)+O\left(\varepsilon^2\ln\varepsilon+\sigma^2\right)\\
&= 2\ds\varepsilon{D e^{-kb}}\left(S g^2+\ds{2\pi\mu}k^{-2} g'^2\right)+O\left(\varepsilon^2\ln\varepsilon+\sigma^2\right).
\end{align*}
Finally, using (\ref{Asec1}), we obtain
$$
\sigma={2\varepsilon^2\;D e^{-kb}} \left( S g^2+{2\pi\mu}{k^{-2}}g'^2\right)+O\left(\varepsilon^3\ln\varepsilon\right).
$$
Note that, since $R_0$ is real and the real parts of all ${M}_{j}$ are even and their imaginary parts are odd, the expression (\ref{EP}) is real for real $\sigma$. Since $F$ is analytic in $\varepsilon, \varepsilon_1$ and $\sigma$, all its Taylor coefficients are real, and hence the solution $\sigma=\sigma(\varepsilon,\varepsilon_1)$ of (\ref{Asec1}) is real up to any order.
Thus we have proven Theorem\;\ref{1}.$\qquad\blacksquare$\\

\subsection{Complex resonance}\label{3.5}
This section, from the technical point of view, is the most complicated. This is due to the fact that the imaginary part of the resonance is very small ($\sim \ve^4$) and hence the perturbative calculations are quite involved.
In this section we will consider a neighborhood of the threshold $\Lambda_2$,
\begin{equation}\label{laa2}
\lambda=\Lambda_2(1-\sigma^2),\quad \sigma\to0\quad\text{as}\quad\varepsilon\to 0.
\end{equation}
For these values of $\lambda$, $\det B$ has four simple zeros in $\mathcal{S}$ at the points $\pm p_1$ and $\pm ip_{0 2}$, where $p_{1}$ is given by (\ref{p^1}) and $p_{0 2}$ by (\ref{p phiphi}).\\

\begin{figure}[!h]
\begin{center}
\includegraphics[width=9cm, height=5.33cm]{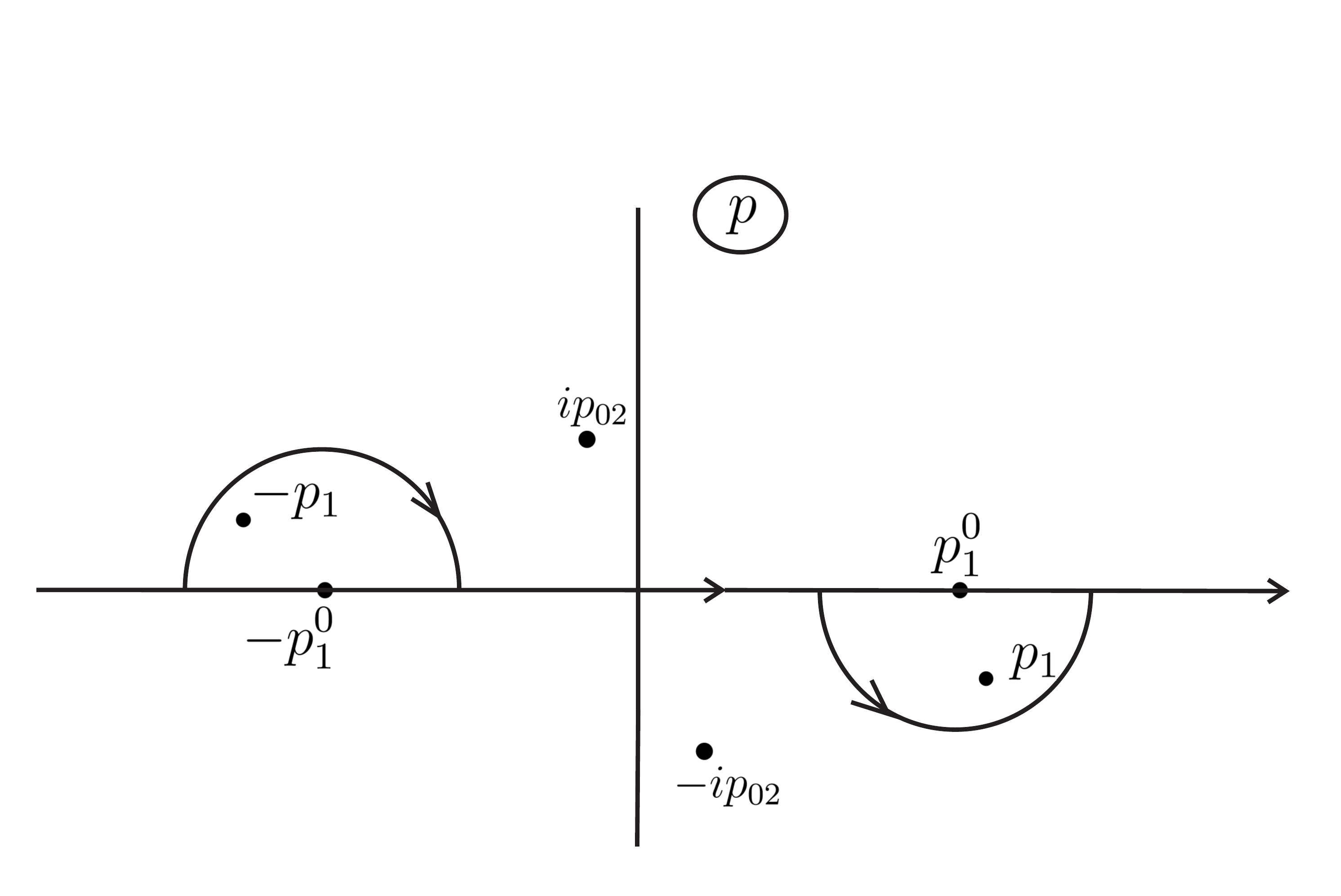}
\end{center}
\caption{Contour\;$\gamma$}  \label{X1}
\end{figure}

In the case under consideration  the solution of the initial problem (\ref{etiqueta_1})-(\ref{2.b.a}) has the form of a slightly modified Fourier transform, namely,
\begin{equation}\label{phips}
\varphi=\ds\frac{1}{2\pi}\ds\int\limits_{\gamma} e^{ipx} \;\tilde{\varphi}(p)\;dp,\quad \psi=\ds\frac{1}{2\pi}\ds\int\limits_{\gamma} e^{ipx} \;\tilde{\psi}(p)\;dp.
\end{equation}
Here the contour $\gamma$ is  shown in Fig.\;\ref{X1} and
 $\tilde{\varphi}$ and $\tilde{\psi}$ satisfy the same system (\ref{ml}) in which the operators $\hat{M}_{2,3}$ have the same kernels but the integration in (\ref{M1}) should be carried out along the contour $\gamma$, as in (\ref{phips}).

We will construct solutions of (\ref{ml}) such that they will be meromorphic in the strip $\mathcal{S}$ (with four simple poles at the points $\pm p_1$, $\pm ip_{02}$) and exponentially decreasing as $|p|\to\infty$. Asymptotic behavior of $\varphi, \psi$ as $|x|\to\infty$ is easily obtained from (\ref{phips}) by means of deforming the contour $\gamma$ into horizontal straight lines lying above (for $x\to+\infty$) or below (for $x\to-\infty$) all the poles. This asymptotic behavior shows that $\varphi$ and $\psi$ given by (\ref{phips}) automatically satisfy the radiation condition, or, in other words, are ``outgoing'' solutions as $|x|\to\infty$.\\
\subsubsection{Exact formula for the resonance}
\begin{figure}[!h]
\begin{center}
\includegraphics[width=14cm, height=5.9cm]{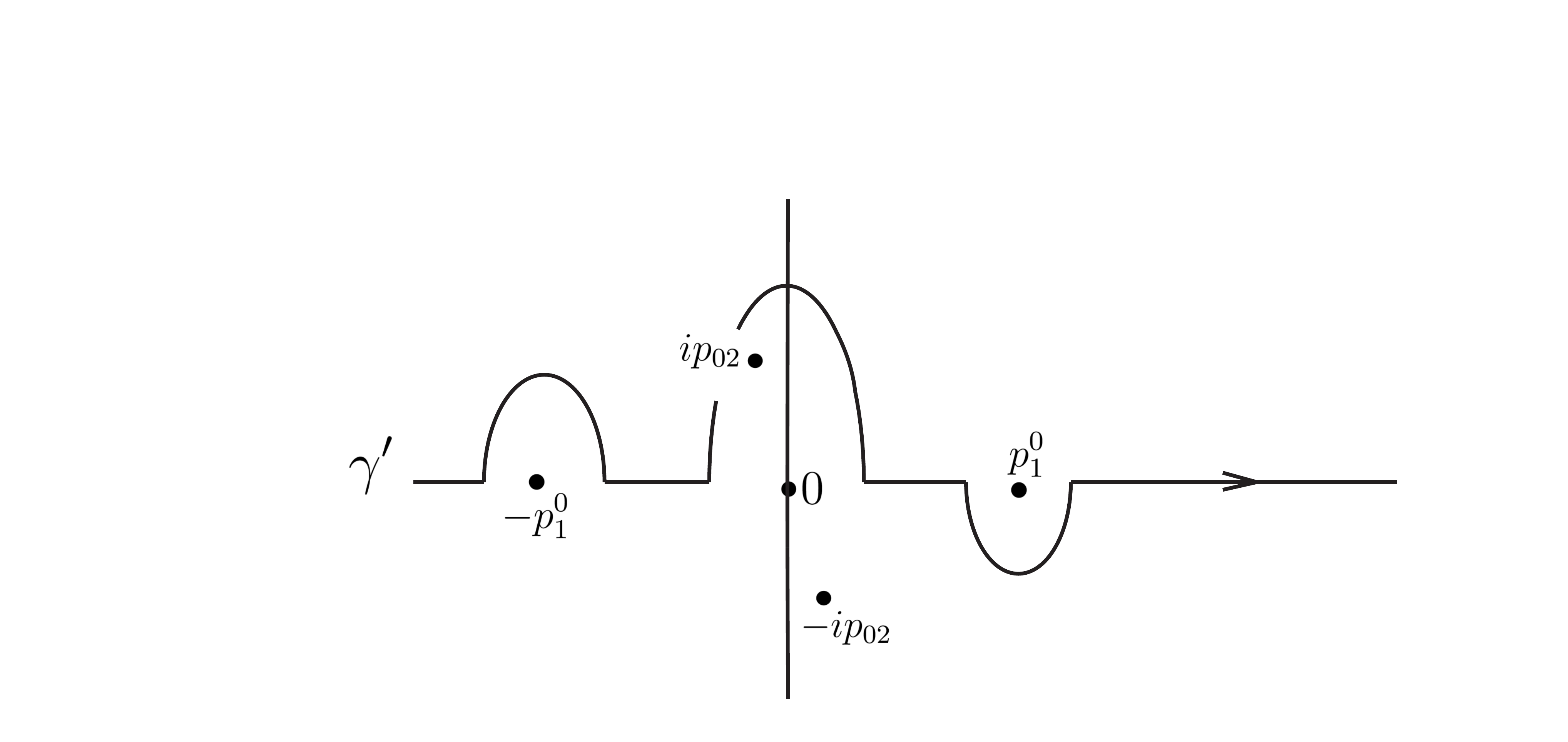}
\end{center}
\caption{Contour $\gamma'$}  \label{TM}
\end{figure}

In this section we obtain the exact secular equation for the resonance which will imply the statement of Theorem \ref{2}.
Just as in the preceding section, equation (\ref{ml}) can be rewritten as an equation for ${\bf A}$:
\begin{equation}\label{eq A}
{\bf A}=\varepsilon \hat{M}\hat{K} {\bf A},
\end{equation}
where
\begin{equation}\label{hat M}
\hat{M}=\begin{pmatrix}
\hat{M}_4\\
\hat{M}_5
\end{pmatrix}\hat{N},\quad \hat{K}{\bf A}=\ds\int\limits_{\gamma}\left(M_2, M_3\right) B^{-1} {\bf A}\;dp,
\end{equation}
and the contour $\gamma$ is shown in Fig.\;\ref{X1}.\\
Since the determinant $\det B$ vanishes at the points $\pm p_1$ and $\pm ip_{02}$, the operator $\hat{M} B^{-1}$ is not necessarily bounded uniformly in $\sigma$ (the points $\pm ip_{02}$ are close to the real axis by (\ref{p phiphi})). Similarly to Lemma \ref{Op Reg}, we have
\begin{lem}\label{Lamb_2}
Let $\lambda=k(1-\sigma^2)$ and let the operator $\hat{K}$ acting from the space $\mathscr{A}^2$ to $C[-\pi,\pi]$ be given by
$$
\hat{K} {\bf A}=\ds\int\limits_{\gamma}\left(M_2(t,p), M_3(t,p)\right) B^{-1} {\bf A}\,dp
$$
where $M_{2,3}$ are given by (\ref{M2}),  $B^{-1}$ is given by (\ref{B-1}).
Then the operator
\begin{equation}\label{KA-}
\hat{T}_0 {\bf A}:=\hat{K} {\bf A}-\ds\frac{1}{\sigma}A_1(0) R_0(t),
\end{equation}
where
\begin{align}\label{RphiE}
R_0(t)&=D ke^{\varepsilon kY},\quad D=\ds\frac{4e^{-ak}}{Q(k) \left(\Lambda_2-\Lambda_1\right)  q_2},
\end{align}
is bounded uniformly in $\sigma$ for small $\sigma$. Here $q_2$ is given by (\ref{p phiphi}).
\end{lem}
{\bf Proof}. The proof, as in Lemma \ref{Op Reg}, consists in changing the contour  $\gamma$ to the one shown in Fig.\ref{TM} and applying the Cauchy residue theorem since the second term in (\ref{KA-}) is simply $2\pi i$ times the residue of the integrand in $\hat{K}{\bf A}$ at the point $ip_{02}$ (note that the expression $\lambda_2'(k)$ appearing in the denominator is equal to 1)  up to terms bounded uniformly in $\sigma$.~~~~$\blacksquare$

\begin{obs}\label{Re}
Note that $R_0(t)$ is real.
\end{obs}
Using (\ref{KA-}), we see that (\ref{eq A}) can be rewritten as the following equation for the function $\bf A$:
\begin{equation}\label{eq.A}
{\bf A}=\varepsilon \hat{T}{\bf A}+\ds\frac{\varepsilon}{\sigma} A_1(0)\hat{M} R_0,
\end{equation}
where
\begin{equation}\label{TMT0}
\hat{T}:=\hat{M}\hat{T}_0.
\end{equation}
Clearly, the operator $\hat{T}$ is bounded in the space $\mathscr{A}^2$ and the solution of (\ref{eq.A}) is given by
\begin{equation}\label{ME}
{\bf A}=\ds\frac{\varepsilon}{\sigma} A_1(0)\left(1-\varepsilon\hat{T}\right)^{-1} \hat{M} R_0.
\end{equation}
We can evaluate $A_1$ at the point $p=0$:
$$
A_1(0)=\ds\frac{\varepsilon}{\sigma} A_1(0)\left.\left(\left(1-\varepsilon\hat{T}\right)^{-1} \hat{M} R_0\right)_{1}\right\vert_{p=0};
$$
here the notation $(\bullet)_j$ means the $j$th component of the vector $\bullet$.\\
Dividing by $A_1(0)$ and multiplying by $\sigma$, we obtain a secular equation for $\sigma$:
\begin{equation}\label{eq.sig}
\sigma=\varepsilon\left.\left(\left(1-\varepsilon\hat{T}\right)^{-1} \hat{M} R_0\right)_{1}\right\vert_{p=0}.
\end{equation}
Clearly, the right-hand side of (\ref{eq.sig}) is analytic in $\sigma, \varepsilon, \varepsilon_1=\varepsilon\ln\varepsilon$. Hence, by the Implicit Function Theorem, we can obtain the solution $\sigma=\sigma(\varepsilon,\varepsilon_1)$ of (\ref{eq.sig}).\\
Just as in the preceding section, ${\bf A}$ is analytic in $\mathcal{S}$ and decreases exponentially as $|p|\to \infty$; thus the pair $\left(\sigma(\varepsilon,\varepsilon_1), {\bf A}\right)$ determines a resonance   for problem $U$ if the imaginary part of $\sigma$ does not vanish. In this way we have obtained an exact formula for the resonance.
\subsubsection{Asymptotics of the real part of the resonance}
To conclude the proof of Theorem\;\ref{2}, we will have to prove formulas (\ref{re-s}) and (\ref{im-s}) for the asymptotics of the real and imaginary parts of $\sigma$. In this subsection we will deal  with the real part.
We have
\begin{equation*}
\left(1-\varepsilon\hat{T}\right)^{-1}\hat{M} R_0=\hat{M} R_0+\varepsilon\hat{T} \hat{M} R_0+O(\varepsilon^2).
\end{equation*}
Thus (\ref{eq.sig}) reads
\begin{equation}\label{Eq sigp}
\sigma=
\left.\left(\varepsilon\left(\hat{M} R_0\right)_{1}+\varepsilon^2\left(\hat{T}\hat{M} R_0\right)_{1}+\varepsilon^3\left(\hat{T}^2\hat{M} R_0\right)_{1}+\cdots\right)\right\vert_{p=0},
\end{equation}
and, by (\ref{RphiE}),
\begin{equation}
 \label{Rphi e^0}
R_0=D k\left(1+\varepsilon k Y+O(\varepsilon^2)\right).
\end{equation}

Let us calculate the asymptotics of the right-hand side of (\ref{Eq sigp}).\\
Taking into account Remark\;\ref{Re}, the fact that $\hat{N}$ sends real-valued functions to real-valued functions  and separating the real and imaginary parts of $\hat{M} R_0$, we obtain
\begin{equation}\label{bf}
\hat{M} R_0=\f_1+i\f_2=\f,
\end{equation}
where, by the definition of $\hat{M}$,
\begin{align*}
\f_1&=\ds\int\limits_{-\pi}^{\pi}
\begin{pmatrix}
\left(-\dot{X}-\dot{X}\varepsilon\tau Y- \ds\frac{\varepsilon p^2}{\tau} X\dot{Y}+O(\varepsilon^2)\right) e^{-a\tau}\\
\left(\dot{X}-\dot{X}\varepsilon\tau Y-\ds\frac{\varepsilon p^2}{\tau} X\dot{Y}+O(\varepsilon^2)\right) e^{-c\tau}
\end{pmatrix}\hat{N} R_0\,dt,\\\\
\f_2&= \ds\int\limits_{-\pi}^{\pi}
\begin{pmatrix}
\left(-\ds\frac{p}{\tau} \dot{Y}+\varepsilon p X\dot{X}-\varepsilon p\ Y\dot{Y}+O(\varepsilon^2)\right) e^{-a\tau}\\
\left(-\ds\frac{p}{\tau} \dot{Y}-\varepsilon p X\dot{X}+\varepsilon p\ Y\dot{Y}+O(\varepsilon^2)\right) e^{-c\tau}
\end{pmatrix}\hat{N} R_0\,dt.
\end{align*}
Note that ${\bf f}_1(p)$ is even and ${\bf f}_2(p)$ is odd.
Also, by (\ref{M_1}), (\ref{NN0}) and (\ref{Rphi e^0}),
\begin{equation}\label{NR0}
\hat{N} R_0=\hat{N}_0 R_0+O\left(\varepsilon^2\ln\varepsilon\right)=Dk\left(\frac{1}{2}+\varepsilon k\hat{N}_0 Y+O\left(\varepsilon^2\ln\varepsilon\right)\right).
\end{equation}
Similarly to Lemma\;\ref{hat{M}R_0 O}, $\hat{M} R_0=O(\varepsilon)$, and, using (\ref{lemma45}) and Proposition\;\ref{NN}, as in (\ref{MR phiphi}), (\ref{MR' phiphi}), we obtain for the first component of $\hat M R_0$
\begin{align}
\nonumber
\left.\left(\hat M R_0\right)_1\right\vert_{p=0}&=-\ds\int\limits_{-\pi}^{\pi}\dot{X} e^{-ak} \left(1+\varepsilon k Y+O(\varepsilon^2)\right) D k\left(\frac{1}{2}+\varepsilon k\hat{N}_0 Y+O\left(\varepsilon^2\ln\varepsilon\right)\right)\,dt\\
\label{MR01}
&=\frac\varepsilon2 D k^2 e^{-ak}\left(S+2\pi\mu\right)+O\left(\varepsilon^2\ln\varepsilon\right).
\end{align}
By (\ref{Eq sigp}), we see that
\begin{equation}\label{sigeps2}
\sigma(\varepsilon, \varepsilon_1)=\frac{\varepsilon^2}2 Dk^2 e^{-ak}\left( S+2\pi\mu\right)+O\left(\varepsilon^3\ln\varepsilon\right),
\end{equation}
and hence $\sigma$ is real in the leading term and thus we have proven formula (\ref{re-s}).
\subsubsection{Imaginary part of the resonance}
 In this subsection we calculate the leading term of the imaginary part of $\sigma$, i.e., we prove formula (\ref{im-s}). In turns out that it is of order of $\varepsilon^4$. Indeed, we have, by (\ref{bf}), $\Im \left.\left(\hat{M} R_0\right)_{1}\right\vert_{p=0}=0$ since ${\bf f}_2$ is odd. Hence, the principal contribution to $\Im \sigma$ comes from the term $\varepsilon^2 (\hat{T}\hat{M} R_0)_1$ in (\ref{Eq sigp}).\\
We have, by the definitions of $\hat T$ (\ref{TMT0}) and $\hat M$ (\ref{hat M}),
\begin{align}\label{TMRphi}
\left.\left(\hat{T}\hat{M} R_0\right)_{1}\right\vert_{p=0}&=\hat{M}_4^0\hat{N}\hat{T}_0 \f,
\end{align}
where
\begin{equation*}
\hat{M}_4^0 h(t):=\ds\int\limits_{-\pi}^{\pi} M_4(0,t) h(t)\,dt=-\ds\int\limits_{-\pi}^{\pi}\dot{X} e^{-(a-\varepsilon Y)k} h(t)\,dt.
\end{equation*}
Since $\hat{M}_4^0$ and $\hat{N}$ send real-valued functions to reals and real-valued functions, respectively, we will be interested in the imaginary part of $\hat{T}_0 \f$.
We have
\begin{equation}\label{Tphif}
\hat{T}_0 \f=I_0-\frac{1}{\sigma}k D e^{\varepsilon k Y} \left(\f_1(0)\right)_{1},
\end{equation}
where $I_0=\int_{\gamma} \left(M_2, M_3\right) B^{-1} \f\,dp$, since $\f_2$ is odd.
Obviously, the second term in (\ref{Tphif}) is real and hence $\Im \hat{T}_0 \f=\Im I_0$.\\

In order to calculate the asymptotics of the term $I_0$ in (\ref{Tphif}), we will use the following statement (the symbol $\ds\fint$ is used for the principal value of the corresponding integrals):
\medskip
\begin{lem}\label{XX}
Let $h(p)$ be analytic in the strip $\mathcal{S}$ and decrease exponentially as $|p|\to \infty$. Let $p_1$ have the form $p_1=p_1^0+p_1^1\sigma^2+O(\sigma^4)$, where $ p_1^0>0$, $ p_1^1<0$, $\Im\sigma>0$, $\Re\sigma>0$. Then
\begin{equation*}
\ds\int\limits_{\gamma_1}\ds\frac{h(p)}{p-p_1}\;dp=\ds\fint\limits_{\R} \ds\frac{h(p)}{p-p_1^0}\;dp-\pi i\; h(p_1^0)+O(\sigma^2).
\end{equation*}
Here the contour $\gamma_1$ is shown in Fig.\;\ref{Agamma_1}.

\begin{figure}[!h]
\begin{center}
\includegraphics[width=9cm, height=5.33cm]{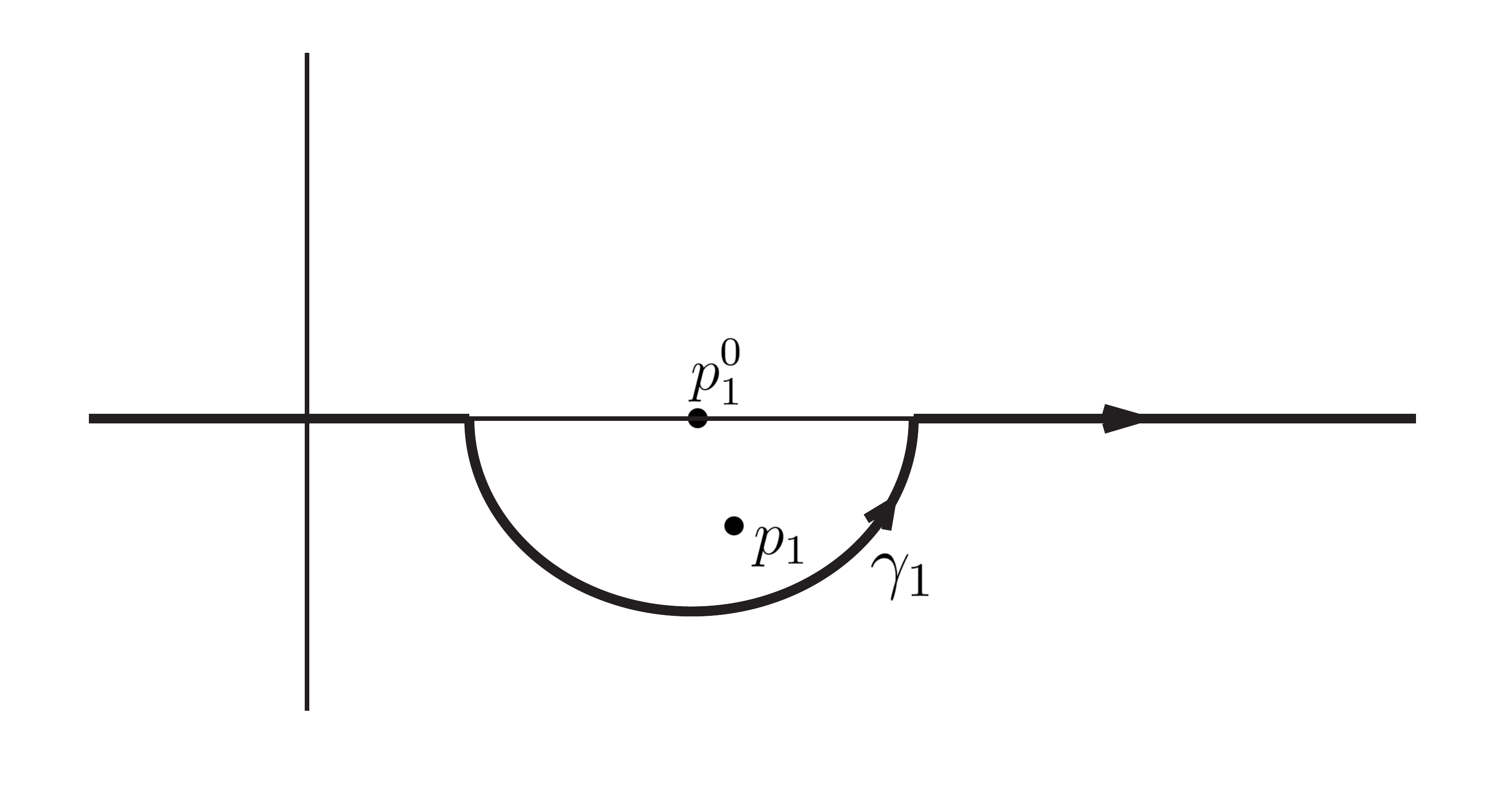}
\end{center}
\caption{Contour $\gamma_1$}  \label{Agamma_1}.
\end{figure}

Similarly,
\begin{equation*}
\ds\int\limits_{\gamma_2}\ds\frac{h(p)}{p+p_1}\;dp=\ds\fint\limits_{\R} \ds\frac{h(p)}{p+p_1^0}\;dp+\pi i\; h(-p_1^0)+O(\sigma^2),
\end{equation*}
where $\gamma_2$ is shown in Fig.\;\ref{Agamma_2}.

\begin{figure}[!h]
\begin{center}
\includegraphics[width=9cm, height=5.33cm]{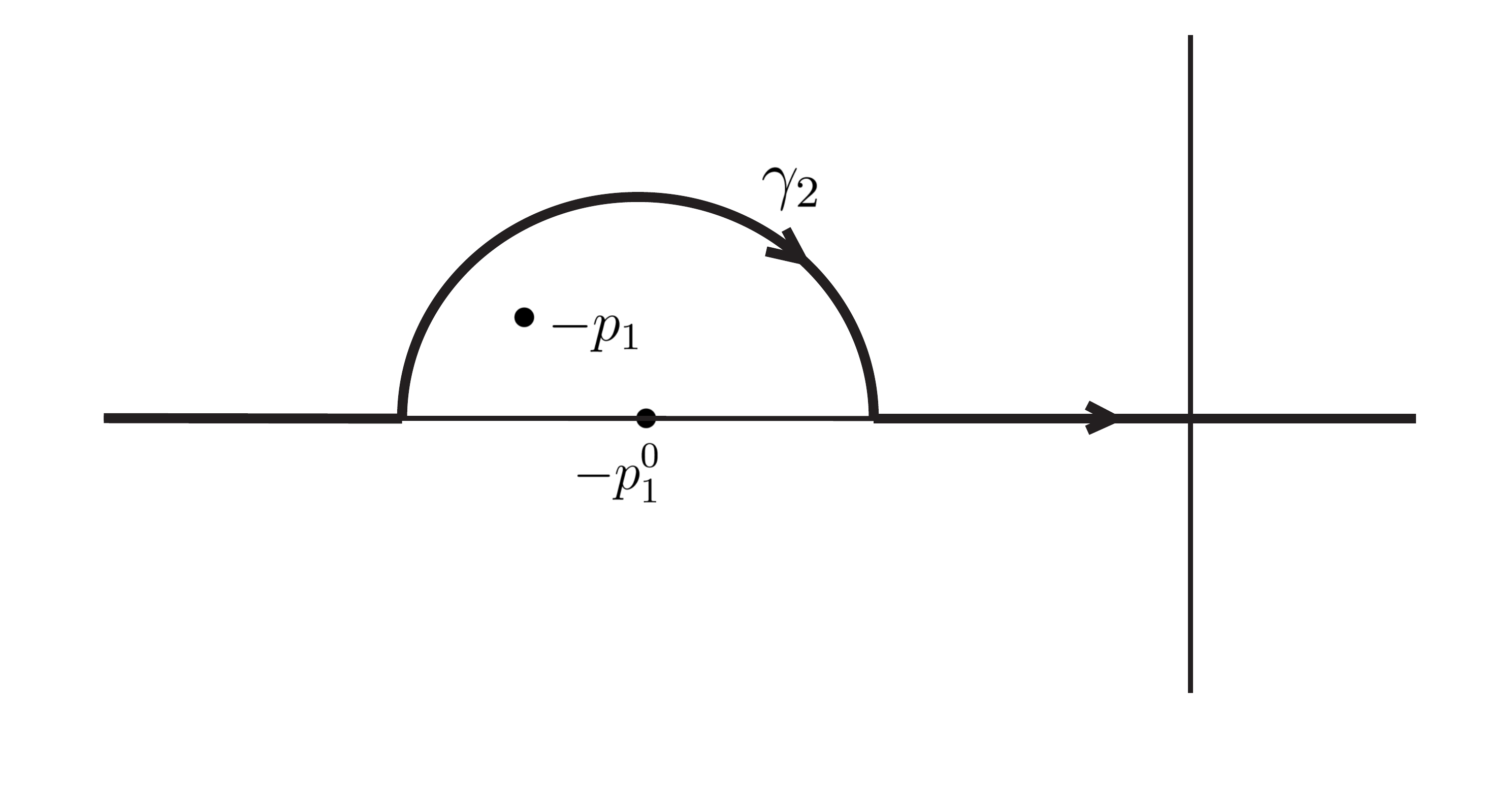}
\end{center}
\caption{Contour $\gamma_2$}  \label{Agamma_2}
\end{figure}
\end{lem}
The proof of this lemma is obvious.~~~~$\blacksquare$

For the integral $I_0$, this lemma implies the following:
\begin{equation}\label{IVP}
I_0= \ds\fint \left(M_2, M_3\right) B^{-1}\f\,dp+\pi i\left.\Res_{p_1}\left(M_2, M_3\right)B^{-1} \f\right|_{\sigma=0}-\pi i\left.\Res_{-p_1}\left(M_2, M_3\right)B^{-1} \f\right|_{\sigma=0}+O(\sigma^2).
\end{equation}
Here in the first summand the factor $(\lambda_1(\tau)-\Lambda_2(1-\sigma^2))$ entering $B^{-1}$ is changed to $(\lambda_1(\tau)-\Lambda_2)$.
At the first step, we show that the  integral in (\ref{IVP}) is real.\\
Let us separate the real and imaginary parts of $M_{j}$ for real $p$. We have, by their definition,
\begin{equation*}
M_{j}=M_{j1}+i M_{j2},\quad j=2,3,4,5,
\end{equation*}
\begin{align}\label{M21}
M_{21}&= \ds\frac{1}{2\pi}\;\ds\frac{\tau+\lambda}{\tau} e^{-a\tau+\varepsilon\tau Y} \cos\varepsilon p X,\quad M_{22}= \ds\frac{1}{2\pi}\;\ds\frac{\tau+\lambda}{\tau} e^{-a\tau+\varepsilon\tau Y} \sin\varepsilon p X,
\end{align}
\begin{align}\label{M31}
M_{31}&= -\ds\frac{1}{2\pi}\;\ds\frac{\tau-\lambda}{\tau}\;\ds\frac{\alpha}{\beta\lambda} e^{-c\tau-\varepsilon\tau Y} \cos\varepsilon p X,\quad M_{32}&= -\ds\frac{1}{2\pi}\;\ds\frac{\tau-\lambda}{\tau}\;\ds\frac{\alpha}{\beta\lambda} e^{-c\tau-\varepsilon\tau Y} \sin\varepsilon p X,
\end{align}
\begin{align}\label{M41}
M_{41}&= e^{-a\tau+\varepsilon\tau Y}\left(-\dot{X}\cos\varepsilon p X-\ds\frac{p}{\tau} \dot{Y} \sin\varepsilon p X\right),\quad M_{42}&= e^{-a\tau+\varepsilon\tau Y}\left(-\ds\frac{p}{\tau} \dot{Y} \cos\varepsilon p X+\dot{X}\sin\varepsilon p X\right),
\end{align}
\begin{align}\label{M51}
M_{51}&= e^{-c\tau-\varepsilon\tau Y}\left(\dot{X}\cos\varepsilon p X-\ds\frac{p}{\tau} \dot{Y} \sin\varepsilon p X\right),\quad M_{52}&= e^{-c\tau-\varepsilon\tau Y}\left(-\ds\frac{p}{\tau} \dot{Y} \cos\varepsilon p X-\dot{X}\sin\varepsilon p X\right).
\end{align}
Obviously, the real parts of all the kernels  $M_{j}$ are even in $p$ and the imaginary parts are odd. We have, since $\f$ has the same property and $B$ is real for real $p$,
\begin{align}\nonumber
&\Im\fint\left(M_{21}+iM_{22}, M_{31}+iM_{32}\right) B^{-1} \left(\f_1+i\f_2\right)\,dp\\\label{Im I}
&\qquad\qquad=\fint\left(M_{21}, M_{31}\right) B^{-1} \f_2\,dp+\fint\left(M_{22}, M_{32}\right) B^{-1} \f_1\,dp=0.
\end{align}
Thus $\Im \hat{T}_0 \f$ comes from the residues in (\ref{IVP}).
Further, by (\ref{IVP}) and (\ref{Im I}),
\begin{equation*}
\begin{array}{lll}
\Im T_0 \f&=&\Im I_0=\Im \pi i\left.\left(\Res_{p_1}\;h-\Res_{-p_1}\;h\right)\right|_{\sigma=0}+O(\sigma^2)\\\\
&=&  \pi \Re\left.\left(\Res_{p_1}\;h-\Res_{-p_1}h\right)\right|_{\sigma=0}+O(\sigma^2),
\end{array}
\end{equation*}
where $h= \left(M_{2}, M_{3}\right) B^{-1}\f$.
Moreover, by parity,
$
\Re\left(\Res_{p_1}h-\Res_{-p_1}h\right)=2\Re \Res_{p_1}h.
$
Thus
\begin{equation*}
\Im I_0=\left.\left(\ds\frac{2\pi}{\left.\left(\frac d{dp}\det B\right)\right\vert_{p_1}}\left.\left[\left(M_{21}, M_{31}\right) J\f_1-\left(M_{22}, M_{32}\right) J\f_2\right]\right|_{p_1}\right)\right\vert_{\sigma=0}+O(\sigma^2),
\end{equation*}
where $J=\operatorname{adj} B$.
We have
$$
\left.J \f\right\vert_{p_1}=\begin{pmatrix}
-\frac{\alpha}{\beta\lambda}\;e^{-b\tau}\\
1
\end{pmatrix} \left.\Ort \f\right\vert_{p_1},
$$
and, analogously to Lemma\;\ref{hat{M}R_0 O},
$
\f_1=O(\varepsilon)$, $\f_2=O(\varepsilon).
$
Let us calculate the leading terms of these expressions. We have
$$
\f_1=\ds\int_{-\pi}^\pi\begin{pmatrix}
M_{41}\\
M_{51}
\end{pmatrix} \hat{N} R_0\,dt,
\qquad
\f_2=\ds\int_{-\pi}^\pi\begin{pmatrix}
M_{42}\\
M_{52}
\end{pmatrix}\hat{N} R_0\,dt.
$$
Using the explicit expressions (\ref{M21}) and (\ref{M31}), we obtain
\begin{equation}\label{JF01}
J \f_1= Dk\begin{pmatrix}
-\ds\frac{\alpha}{\beta\lambda}e^{-b\tau}\\
1
\end{pmatrix}
 I_1,
\end{equation}
where
\begin{align}\nonumber
I_1&= \frac{\tau+\lambda}{\tau}e^{-b\tau-a\tau} \int_{-\pi}^\pi e^{\varepsilon\tau Y}\left(-\dot{X}\cos p\varepsilon X-\ds\frac{p}{\tau} \dot{Y}\sin p\varepsilon X\right)\hat{N} e^{\varepsilon k Y}\,dt\\\label{I_1}
&\quad+  \frac{\tau-\lambda}{\tau} e^{-b\tau+a\tau} \int_{-\pi}^\pi e^{-\varepsilon \tau Y}\left(\dot{X}\cos p\varepsilon X-\ds\frac{p}{\tau} \dot{Y}\sin p\varepsilon X\right) \hat{N} e^{\varepsilon k Y}\,dt.
\end{align}
Hence, by (\ref{M21}), (\ref{M31}),
\begin{equation}\label{JF1}
\left(M_{21}, M_{31}\right) J \f_1=-\frac{1}{2\pi}\frac{\alpha Dk}{\beta\lambda\tau}e^{-b\tau}\left(({\tau+\lambda})e^{-a\tau+\varepsilon Y \tau}+(\tau-\lambda) e^{a\tau-\varepsilon Y\tau}\right)(\cos p\varepsilon X)I_1.
\end{equation}
Similarly,
\begin{equation}\label{JF02}
J \f_2= Dk\begin{pmatrix}
-\ds\frac{\alpha}{\beta\lambda}e^{-b\tau}\\
1
\end{pmatrix} I_2,
\end{equation}
where
\begin{align}\nonumber
I_2&= \frac{\tau+\lambda}{\tau}e^{-b\tau-a\tau}\int_{-\pi}^\pi e^{\varepsilon\tau Y}\left(-\frac{p}{\tau} \dot{Y}\cos p\varepsilon X+\dot{X}\sin p\varepsilon X\right)\hat{N} e^{\varepsilon k Y}\;dt\\\label{I_2}
&\quad+\frac{\tau-\lambda}{\tau} e^{-b\tau+a\tau} \int_{-\pi}^\pi e^{-\varepsilon \tau Y}\left(-\frac{p}{\tau} \dot{Y}\cos p\varepsilon X-\dot{X}\sin p\varepsilon X\right) \hat{N} e^{\varepsilon k Y}\,dt.
\end{align}
Hence, by (\ref{M21}), (\ref{M31}),
\begin{equation}\label{JF2}
\left(M_{22}, M_{32}\right) J \f_2=-\frac{1}{2\pi}\frac{\alpha Dk}{\beta\lambda\tau}e^{-b\tau}\left(({\tau+\lambda})e^{-a\tau+\varepsilon Y \tau}+(\tau-\lambda) e^{a\tau-\varepsilon Y\tau}\right)(\sin p\varepsilon X)
 I_2.
\end{equation}
Let us calculate the asymptotics of integrals entering $I_1$. We have, similarly to (\ref{MR phiphi}), (\ref{MR' phiphi}), and recalling that $\sigma=O(\ve^2)$ by (\ref{sigeps2}),
\begin{equation}\label{In+1}
\int\limits_{-\pi}^{\pi} e^{\varepsilon\tau Y}\left(-\dot{X} \cos p\varepsilon X-\ds\frac{p}{\tau} \dot{Y}\sin p\varepsilon X\right) \hat{N} e^{\varepsilon k Y}\,dt
=\frac{\varepsilon}{2}k\left(\frac{k}{\tau}S+2 \pi\mu\right)+O\left(\varepsilon^2\ln\varepsilon\right).
\end{equation}
Similarly,
\begin{equation}\label{In+2}
\int\limits_{-\pi}^{\pi} e^{-\varepsilon\tau Y}\left(\dot{X} \cos p\varepsilon X-\frac{p}{\tau} \dot{Y}\sin p\varepsilon X\right)\hat{N} e^{\varepsilon k Y}\,dt
=\frac{\varepsilon}{2}k\left(\frac{k}{\tau}S-2 \pi\mu\right)+O\left(\varepsilon^2\ln\varepsilon\right).
\end{equation}
Substituting in (\ref{I_1}), we have
\begin{align}\nonumber
 I_1 &= \frac{\tau+\lambda}{\tau}e^{-b\tau-a\tau}\frac{\varepsilon}{2}\left(\frac{k^2}{\tau}S+2k\pi\mu\right)+
\frac{\tau-\lambda}{\tau}e^{-c\tau}\frac{\varepsilon}{2}
\left(\ds\frac{k^2}{\tau}S-2k\pi\mu\right)+O\left(\varepsilon^2\ln\varepsilon\right)\\\label{{I_1}}
&=\frac{ e^{-b\tau}}{\tau^2}k\varepsilon\left(kSg\left(-a,\tau,\lambda\right)+2\pi\mu g'\left(-a,\tau,\lambda\right)\right)+O\left(\varepsilon^2\ln\varepsilon\right).
\end{align}
Now we calculate integrals entering $I_2$. We have, similarly to the above,
\begin{equation}\label{In+3}
\int\limits_{-\pi}^{\pi} e^{\varepsilon\tau Y}\left(-\frac{p}{\tau} \dot{Y}\cos p\varepsilon X+\dot{X} \sin p\varepsilon X\right)\hat{N} e^{\varepsilon k Y}\,dt
=-\varepsilon\ds\frac{pk}{\tau}\;\pi\nu+O\left(\varepsilon^2\ln\varepsilon\right).
\end{equation}
In the same way,
\begin{equation*}
\int\limits_{-\pi}^{\pi} e^{-\varepsilon\tau Y}\left(-\frac{p}{\tau} \dot{Y}\cos p\varepsilon X-\dot{X} \sin p\varepsilon X\right)\hat{N} e^{\varepsilon k Y}\,dt
=-\varepsilon\ds\frac{p k}{\tau} \pi\nu+O\left(\varepsilon^2\ln\varepsilon\right).
\end{equation*}
Substituting in (\ref{I_2}), we have
\begin{align}
\nonumber
 I_2 &= -\varepsilon\frac{p k}{\tau^2}\;\pi\nu\left[\left(\tau+\lambda\right) e^{-b\tau-a\tau}+\left(\tau-\lambda\right) e^{-c\tau}\right]+O\left(\varepsilon^2\ln\varepsilon\right)\\\label{{I_2}}
 &=-2\varepsilon\frac{p k}{\tau^2}e^{-b\tau}\pi\nu g(-a,\tau,\lambda)+O\left(\varepsilon^2\ln\varepsilon\right).
\end{align}
Recall that, by (\ref{IVP}), we have for the imaginary part of (\ref{TMRphi})
\begin{align}
\nonumber
\mathcal{F}:&=\Im\hat{M}_4^0\;\hat{N}\hat{T}_0\hat{M} R_0\\\label{FCur'}&=-2\pi D_1\hat{M}_4^0\hat{N}\left.\left[\left(M_{21},M_{31}\right) J \f_1-
\left(M_{22},M_{32}\right) J \f_2\right]\right\vert_{p_1^0}+O(\sigma^2),\end{align}
 where
 \begin{equation}\label{FCur}D_1=\left.\left(\frac{-1}{\left.\left(\frac d{dp}\det B\right)\right\vert_{p_1}}\right)\right|_{\sigma=0}=\ds\frac{\Lambda_2\tau_1}{Q(\tau_1)\; \lambda'_1(\tau_1)p_1^0 \left(\lambda_2(\tau_1)-\Lambda_2\right)}.
\end{equation}
Expanding  expressions (\ref{JF1}), (\ref{JF2})  in $\varepsilon$, and recalling (\ref{JF01}), (\ref{JF02}), we have in terms of $I_{1,2}$
\begin{equation*}
\left(M_{21},M_{31}\right) J\f_1=-\frac{1}{2\pi}\frac{\alpha Dk}{\beta\lambda\tau}e^{-b\tau}\left((\tau+\lambda)e^{-a\tau}\left(1+\varepsilon Y \tau\right) +(\tau-\lambda)e^{a\tau}\left(1-\varepsilon Y \tau\right) \right)  I_1+O(\varepsilon^3),
\end{equation*}
\begin{equation*}
\left(M_{22},M_{32}\right) J\f_2=-\frac{1}{2\pi}\frac{\alpha Dk}{\beta\lambda\tau}e^{-b\tau}\left((\tau+\lambda)e^{-a\tau} +(\tau-\lambda)e^{a\tau} \right)\varepsilon p X Dk I_2+O(\varepsilon^3).
\end{equation*}
Substituting in (\ref{FCur'}) and performing elementary calculations as in (\ref{MR phiphi}), (\ref{MR' phiphi}), we have
\begin{align*}
\mathcal{F}&=-\frac{\alpha}{\beta\lambda\tau}DD_1ke^{-b\tau-ak}\ds\int_{-\pi}^\pi\dot{X}\left(1+\varepsilon k Y+O(\varepsilon^2)\right)\left(\hat{N}_0+O\left(\varepsilon^2\ln\varepsilon\right)\right)\\
&\qquad\times\Big[\left((\tau+\lambda)e^{-a\tau} \left(1+\varepsilon Y\tau\right)+({\tau-\lambda})e^{a\tau} \left(1-\varepsilon Y\tau\right)\right) I_1\\
&\qquad\qquad- \left((\tau+\lambda)e^{-a\tau}+(\tau-\lambda)e^{a\tau}\right)\varepsilon p X I_2\Big]\Big|_{\sigma=0,p=p_1^0}\,dt+O(\varepsilon^3+\sigma^2).
\end{align*}
Substituting here (\ref{{I_1}}), (\ref{{I_2}}) instead of $I_{1,2}$, and using  formula (\ref{nu1}) we obtain, similarly to (\ref{In+1}), (\ref{In+2}), (\ref{In+3}),
\begin{equation}
\mathcal{F}= \ds\frac{e^{-ak-2b\tau_1}}{\tau_1^3}k^2DD_1\ds\frac{\alpha}{\beta\Lambda_2}\varepsilon^2 \left\lbrace \left(k Sg+
 2\pi\mu g'\right)^2\label{FC}
 + \left(2p_1^0g\pi\nu\right)^2\right\rbrace+O\left(\varepsilon^3\ln\varepsilon+\sigma^2\right);
\end{equation}
here the arguments $(-a;\tau_1,\Lambda_2)$ of $g$ and $g'$ are omitted.

It is easy to see, analogously to Lemma\;\ref{hat{M}R_0 O}, that
$
\hat{T} \f=O\left(\varepsilon^2\ln\varepsilon\right)
$,
and hence
\begin{equation}\label{third}
\varepsilon^3\hat{T}^{2} \f=O\left(\varepsilon^5\ln\varepsilon\right);
\end{equation}
thus only the term $\varepsilon^2\hat{T} f$ in (\ref{Eq sigp}) contributes to $\Im \sigma$, up to $O(\varepsilon^4)$.
Recalling that $\Im \sigma=\varepsilon^2 \mathcal{F}$ by (\ref{third}) and that $\sigma^2=O(\varepsilon^4)$, we obtain formula (\ref{im-s}) and thus complete the proof of  Theorem\;\ref{2}.

\subsection{Embedded eigenvalue}\label{3.6}
We will prove here Theorem \ref{LAN3}, our central result. As in the previous subsection, we work in a neighborhood of the threshold $\Lambda_2$ assuming that $\lambda$ satisfies (\ref{laa2}).  If the solution of (\ref{ml}) has poles in the strip $\mathcal{S}$ only at the points $\pm ip_{02}$ (see(\ref{p phiphi})), then the corresponding functions $\varphi, \psi$ describe a trapped mode, as in section \ref{3.4}. By (\ref{phi psi}), $\tilde\varphi,\tilde\psi$ are analytic at  $\pm p_1$ if
 the vector ${\bf A}=B(\tilde{\varphi}, \tilde{\psi})^\top$ at the points $p=\pm p_1$ is orthogonal to the $\operatorname{coker}B(\tau_1^*)$ (since $\det B=0$ at these points, the $\operatorname{coker}B$ is nontrivial; we understand here $B(\tau_{1}^{\ast})$ as an operator on $\C^2$; $\tau_1^*$ given by (\ref{tansig}) is the solution of (\ref{lam1}) coinciding with $\tau_1$ for $\sigma=0$). By (\ref{B}) and the definition of $p_1$, the $\operatorname{coker}B$ is spanned by the complex conjugate of the vector $\left.\left((\tau+\lambda) e^{-b\tau}, \tau-\lambda\right)\right|_{\pm p_1}$. Hence the orthogonality conditions, which are sufficient  for the analyticity of $\tilde\varphi,\tilde\psi$  at  $\pm p_1$, read (up to a multiplication by $\tau\neq0$; see (\ref{Ort}))
\begin{equation}\label{orc}
\left.\left(\Ort {\bf A}\right)\right\vert_{p=\pm p_1}=0.
\end{equation}
 By (\ref{ME}), they have the form
\begin{equation}\label{ort c}
\Ort\left.\left(\left(1-\varepsilon\hat{T}\right)^{-1}\hat{M} R_0\right)\right\vert_{p=\pm p_1}=0.
\end{equation}
In the leading term we have
\begin{equation*}
\left.\Ort\left(\hat M R_0\right)\right|_{\pm p_1}=0.
\end{equation*}
Using (\ref{lemma45}), Proposition\;\ref{NN}, and performing calculations similar to (\ref{Rphi e^0}), (\ref{NR0}), (\ref{MR01}), we obtain
\begin{align*}
\left.\Ort\left(\hat M R_0\right)\right|_{\pm p_1}&= \varepsilon D e^{-b\tau}\Big\lbrace 2\pi k\mu\left(\lambda\cosh a\tau-\tau\sinh a\tau\right)+ \ds\frac{k^2}{\tau}S\left(\tau\cosh a\tau-\lambda\sinh a\tau\right)\\
&\qquad\qquad\quad- 2i\ds\frac{p}{\tau}\pi k\nu\left(\tau\cosh a\tau-\lambda\sinh a\tau\right)\Big\rbrace\Big\vert_{\pm p_1}+O\left(\varepsilon^2\ln\varepsilon\right)\\
&=\left.\varepsilon {D e^{-b\tau} \frac k\tau}\left\lbrace k Sg+2\pi\mu g'-2i p\pi\nu g\right\rbrace\right|_{\pm p_1}+O\left(\varepsilon^2\ln\varepsilon\right);
\end{align*}
here  the  arguments $(-a;\tau,\lambda)$ of $g$ and $g'$ are omitted.
Assume that the contour $C$ is symmetric with respect to the $y$-axis. Then $\nu=0$ and the last expression vanishes in the leading term if (cf. (\ref{mathcal{R}}))
\begin{equation}\label{calR}
\mathcal{R}=
 kSg(-a;\tau_1,\Lambda_2)+2\pi\mu g'(-a;\tau_1,\Lambda_2)=0.
\end{equation}
This equation can be considered as an equation for $a$. Recall that $\tau_1$ is the positive root of $\lambda_1(\tau)=\Lambda_2$. Let us rewrite this equation in terms of the dimensionless variables
\begin{equation}\label{orlt}
\tau_0=\frac{\tau_1}k=\sqrt{1+\ds\frac{(p_1^0)^2}{k^2}},\quad a_0=ka, \quad\text{and}\quad b_0=kb.
\end{equation}
 We have, denoting $w=a_0\tau_0$, dividing by $2\pi\mu kg$ and recalling the form of $g$, $g'$,
\begin{equation}\label{fr1}
\ds\frac{S}{2\pi\mu}=\ds\frac{\tau_0^2\sinh w-\tau_0\cosh w}{\tau_0\cosh w-\sinh w}.
\end{equation}
Note that here, since $\lambda_1(\tau_1)=\Lambda_2$, $\tau_0$ is the solution of
\begin{equation}\label{fr2}
1=\ds\frac{\alpha\tau_0\tanh b_0\tau_0}{1+\beta\tanh b_0\tau_0}.
\end{equation}
Equation (\ref{fr1}) is an equation for $w$.
Denote $\delta=\ds\frac{S}{2\pi\mu}$; we have $0<\delta<1$ by (\ref{mu}). Then (\ref{fr1}) can be rewritten as
\begin{equation}\label{subw}
\tanh w=\ds\frac{\tau_0(1+\delta)}{\tau_0^2+\delta}.
\end{equation}
This equation always possesses a solution if $\tau_0$ is sufficiently large (this will be the case, for example, if $\alpha$ in (\ref{fr2}) is sufficiently small, that is, if the densities of the layers are close to each other). On the other hand, $a_0={w}/{\tau_0}$ should satisfy $a_0<b_0$. Again, if $\tau_0$ is sufficiently large, this inequality is satisfied.  Let us assume that $\tau_0$ is such that (\ref{subw}) possesses a solution satisfying the inequality $w/\tau_0<b_0$. We see that (\ref{orc}) in the leading term is satisfied if the submergence of the cylinder satisfies (\ref{subw}); the value $a^*$ mentioned in Theorem \ref{LAN3} is given by $a^*=a_0/k=w/(k\tau_0)$, where $w$ solves (\ref{subw}). This follows directly from the fact that (\ref{subw}) is equivalent to (\ref{calR}).

As mentioned above, $a^*$ is always less than $b$ for small values of $\alpha$. In more detail, for small $\alpha$ we have $\tau_0\sim2/\alpha$ by (\ref{fr2}) and (\ref{subw}) gives $w\sim\alpha(1+\delta)/2$. Hence $a_0=w/\tau_0\sim\alpha^2(1+\delta)/4$ and $a^*=a_0/k\sim\alpha^2(1+\delta)/4k$. The last expression, obviously, is less than $b$ for sufficiently small $\alpha$. Thus an embedded trapped mode always exists for small $\alpha$. The submergence $a^*$ which guarantees its existence must be quite small.

On the other hand, if $\alpha$ is close to 1 ($\beta\to0$), the solution $\tau_0$ of (\ref{fr2}) coincides (up to $O(\beta)$) with the solution $\hat\tau_0$ of $1=\tau_0\tanh b_0\tau_0$; $\tau_0=\hat\tau_0+O(\beta)$. Clearly, $\hat\tau_0=1+\delta_1$ with a positive $\delta_1$ independent of $\beta$. The solution $w=a_0\tau_0$ of (\ref{subw}) must satisfy $w/\tau_0<b_0$. This means that
$$
\tanh a_0\tau_0=\frac{\tau_0(1+\delta)}{\tau_0^2+\delta}<\tanh b_0\tau_0
$$
by the monotonicity of $\tanh$. But, since $\tau_0=\hat\tau_0+O(\beta)$, this means that
$$\frac{\hat\tau_0(1+\delta)}{\hat{\tau}_0^2+\delta}<\frac1{\hat\tau_0}+O(\beta)\qquad\text{or}\qquad \hat\tau_0^2<1+O(\beta),
$$
and the last inequality cannot be true for sufficiently small $\beta$ since $\hat\tau_0=1+\delta_1$, $\delta_1>0$. Thus for sufficiently small $\beta$ there are no embedded trapped modes.

Now, having established sufficient conditions for the existence/nonexistence of solutions of (\ref{calR}), we proceed with the proof of Theorem \ref{LAN3}.
In order to satisfy (\ref{ort c}) up to any order in $\varepsilon$ and $\sigma$, and thus prove the existence of a real value of the parameter $a$ such that there exists a trapped mode, we will investigate the parity properties of the operators entering (\ref{ME}) and (\ref{ort c}). Using these properties, we will prove that equation (\ref{ort c}) is real and hence its solution is also real.

Assume that the lower point of intersection of the contour $C$ with the vertical axis corresponds to the value $t=0$. Then, by symmetry, $X(t)$ is odd and $Y(t)$ is even.
By the explicit formula for the operator $\hat M_1$ (\ref{M_1the}), it is easy to see that $\hat{M}_1$ preserves the parity, that is, $\hat{M}_1 \theta$ is even if $\theta$ is even and odd if $\theta$ is odd. Hence $\Big(1+\hat{M}_1\Big)^{-1}$ has the same property. By (\ref{RphiE}), $R_0$ is even in $t$, hence $\hat N R_0$ is also even in $t$. By (\ref{M41}) and (\ref{M51}), $M_{42}(p,t)$ and $M_{52}(p,t)$ are odd in $t$. Hence,
$
\Im \left(\hat{M} R_0\right)=0
$
on the real axis and for real values of  parameters, and $\hat{M} R_0$ is even in $p$ since $M_{41}$ and $M_{51}$ are even in $p$ and $t$.
Consider equation (\ref{ME}):
\begin{equation}\label{ME1}
{\bf A}=\ds\frac{\varepsilon}{\sigma}\left(1-\varepsilon\hat{T}\right)^{-1}\hat{M} R_0
\end{equation}
(we can set $A_1(0)=1$ without loss of generality). Since $\hat MR_0$ is even in $p$, we obtain by induction that ${\bf A}(p)$ is also even. We can assume in the equivalent  equation (\ref{eq.A}) that the integration in $\hat{K} {\bf A}$ is carried out along the real axis by (\ref{orc}) (the zeros of $\det B$ are located exactly at the points $\pm p_1$ and (\ref{orc}) guarantees that $B^{-1}{\bf A}$ does not have poles at these points). We have
\begin{equation}\label{TA}
\hat{T}{\bf A}
= \hat{M} \ds\int\Big(M_2,M_3\Big) B^{-1}{\bf A}\;dp-\ds\frac{1}{\sigma}\hat{M} R_0.
\end{equation}
The integral operator in the first summand of (\ref{TA}) preserves parity in $p$, and, since ${\bf A}(p)$ is even in $p$ by (\ref{ME1}),  the imaginary part of the first summand in (\ref{TA}) vanishes on the real axis for even real-valued on the real axis ${\bf A}$ which satisfies (\ref{orc}). Hence ${\bf A}$ satisfies, on the real axis and for real values of parameters,  a real equation (since $\hat MR_0$ is real on the real axis) and is itself real. This means that the solution $a=a(\sigma,\varepsilon)$ of (\ref{orc}) is real for real $\sigma$; it exists by the Implicit Function Theorem, for example, for sufficiently large $\tau_0$ since, as it is easy to see, for such $\tau_0$ the derivative of  $\mathcal{R}$ from (\ref{calR}) with respect to $a$ does not vanish. The solution of (\ref{eq.sig}) is also real by the same argument. Theorem \ref{LAN3} is proven. $\qquad\blacksquare$

\section{Cylinder in the lower layer}
\setcounter{equation}{0}

\subsection{System of integral equations}\label{4.1}

Introduce the functions $\varphi, \theta$ by the formulas
\begin{equation*}
\vf:=\left.\phi_1\right\vert_{y=0},\qquad \theta:=\left.\phi_2\right\vert_{\Gamma}.
\end{equation*}
As in section\;3.1, if these functions are known, then $\phi_1$ and $\phi_2$ can be reconstructed as  solutions of (\ref{etiqueta_1}), (\ref{etiqueta_1_2}). Indeed, the conditions $\left.\phi_1\right|_{y=0}=\varphi$ and $\left.\phi_{1y}\right|_{y=0}=\lambda\varphi$ (see (\ref{etiqueta_2_2})) define $\phi_1$ in $\Omega_1$ by means of the Fourier transform since
for the Fourier transform $\tilde{\phi}_1(p,y)$ we have
\begin{equation*}
\left\{  \begin{array}{rcl}
\tilde{\phi}_{1 yy}-\tau^2\tilde{\phi}_1=0,\quad -b<y<0,\\\\
\tilde{\phi}_{1 y}=\lambda\tilde{\phi}_{1},\quad \tilde{\phi}_1=\tilde{\varphi},\quad y=0.
\end{array}
\right.
\end{equation*}
Hence we have
\begin{equation}\label{phi1s}
\tilde{\phi}_{1}=\tilde{\varphi}(p)\left(\tau\cosh \tau y+\lambda\sinh \tau y\right)/\tau
\end{equation}
and $\phi_1$ is completely defined by $\varphi$. Further (see Appendix 2),
\begin{equation}\label{psi1}
\tilde{\psi}:=\left.\tilde{\phi}_{1y}\right\vert_{y=-b}
=\tilde{\varphi}\left(\lambda-\tau\tanh\tau b\right)\cosh\tau b
\end{equation}
and, by (\ref{etiqueta_2_3}), (\ref{etiqueta_2_4}), and since $\alpha=1-\beta$
\begin{equation}\label{Phi2}
\tilde{\phi}_{2}(p,-b)= \ds\frac{1}{\lambda}\alpha\tilde{\psi}(p)+\beta\tilde{\phi}_1(p,-b)=\ds\frac{1}{\lambda\tau}\tilde{\varphi}(p)
\left(\lambda\tau-\left(\alpha\tau^2+\beta\lambda^2\right)\tanh \tau b\right)\cosh\tau b.
\end{equation}
We see that the normal derivative of $\phi_2$ and its value on $\Gamma_I$ are known if $\varphi$ is known; its normal derivative and its value on $\Gamma$ are $0$ and $\theta$, respectively. Thus $\phi_1$ is given by (\ref{phi1s}) and $\phi_2$, by (\ref{GF1}), if $\varphi$ and $\theta$ are known. These latter functions, as shown in Appendix 2, satisfy  equations
(\ref{tilde{phi}_2}) and (\ref{hat{M}}). Substituting  expressions (\ref{psi1}) and (\ref{Phi2}) in (\ref{tilde{phi}_2}) and (\ref{hat{M}}),   we come to the following system for $\tilde{\varphi}$, $\theta$:
\begin{equation*}
\left(1+\hat{M}_1\right)\theta=\ds\frac{1}{2\pi}\ds\int e^{-(a-\varepsilon Y)\tau+ip\varepsilon X} P^+(\tau,\lambda)\tilde{\varphi}(p)\,dp,
\end{equation*}
\begin{equation*}
P^-(\tau,\lambda)\tilde{\varphi}=-\varepsilon\ds\int\limits_{-\pi}^{\pi}\left(\ds\frac{ip}{\tau}\dot{Y}+\dot{X}\right) e^{-(a-\varepsilon Y)\tau-ip\varepsilon X}\theta(t)\,dt,
\end{equation*}
where the coefficients $P^\pm$ are given by
$$P^\pm(\tau,\lambda)=\left(\left(1-\left(\alpha\ds\frac{\tau}{\lambda}+\beta\ds\frac{\lambda}{\tau}\right)\tanh\tau b\right)\pm\ds\frac{1}{\tau}\left(\lambda-\tau\tanh\tau b\right)\right)\cosh\tau b
$$
and $\hat M_1$ is defined in (\ref{M_1the}).
After some algebraic manipulations, we come to the following system:
\begin{equation}\label{theta2}
\left(1+\hat{M}_1\right)\theta=\frac1\lambda\ds\int L_1(t,p)P(\tau)\tilde{\varphi}\,dp,
\end{equation}
\begin{equation}\label{phi2}
\frac1\lambda P(\tau)\left(\lambda_1(\tau)-\lambda\right)\left(\lambda_2(\tau)-\lambda\right)\tilde{\varphi}=\varepsilon\ds\int L_2(p,t)\theta(t)\,dt,
\end{equation}
where
$$
L_1(t,p)=\ds\frac{1}{2\pi}P_0(\tau,\lambda) e^{-(a-\varepsilon Y)\tau+ip\varepsilon X}, \quad L_2(p,t)=\left(\ds\frac{ip}{\tau}\dot{Y}+\dot{X}\right)e^{-(a-\varepsilon Y)\tau-ip\varepsilon X},
$$
\begin{equation}\label{P0}
P_0(\tau,\lambda)=\ds\frac{1-\beta\tanh\tau b}{1+\beta\tanh \tau b}\left(\lambda+\tau\right) \left(\lambda-\ds\frac{\alpha\tau\tanh\tau b}{1-\beta\tanh\tau b}\right),
\end{equation}
$$
P(\tau)=\frac1{\tau}\left(\cosh\tau b+\beta\sinh\tau b\right).
$$
Note that
$$
P^+(\tau,\lambda)=\frac1\lambda P(\tau)P_0(\tau,\lambda)\quad\text{and}\quad P^-(\tau,\lambda)=\frac1\lambda P(\tau)(\lambda_1(\tau)-\lambda)(\lambda_2(\tau)-\lambda).
$$
As in section \ref{3.4}, the operator $(1+\hat{M}_1)$ is invertible and hence (\ref{theta2}) yields (see (\ref{NN0}))
\begin{equation*}
\theta=\frac1\lambda\hat{N} \hat L_1\left(P(\tau)\tilde{\varphi}(p)\right).
\end{equation*}
Substituting in (\ref{phi2}), we come to a single equation for $\tilde{\varphi}$:
\begin{equation}\label{phi 3}
P(p)\left(\lambda_1(\tau)-\lambda\right) \left(\lambda_2(\tau)-\lambda\right)\tilde{\varphi}=\varepsilon\hat L_2\hat{N}\hat L_1\left( P\tilde{\varphi}\right).
\end{equation}
The structure of zeros of the factors $\lambda_1-\lambda$ and $\lambda_2-\lambda$ was already explained in section\;3.3.

\subsection{Discrete eigenvalue}\label{4.2}
In this section we prove Theorem \ref{LN4} for the discrete eigenvalue for the cylinder in the lower layer.
In this case, as in section \ref{3.4}, we assume that $\lambda=\Lambda_1(1-\sigma^2)$ and look for the solution of (\ref{phi 3}) in the form
\begin{equation}\label{subst p}
\tilde{\varphi}=\ds\frac{A(p)}{\left(\lambda_1(\tau)-\lambda\right) \left(\lambda_2(\tau)-\lambda\right) P(\tau)},\quad A\in\mathscr{A},
\end{equation}
with $A(p)$ a new unknown (scalar) function. Substituting in (\ref{phi 3}) we obtain
\begin{equation}\label{All}
A=\varepsilon \hat L_2 \hat{N} \hat L_1\;\ds\frac{A}{\left(\lambda_1-\lambda\right) \left(\lambda_2-\lambda\right)}.
\end{equation}
Exactly as in section \ref{3.4}, we have the following
\begin{lem}\label{XX_x}
The operator
\begin{equation*}
\hat{T}_{0} A=\hat{K} A-\ds\frac{1}{\sigma}A(0)R_0(t)
\end{equation*}
acting from $\mathscr{A}$ to $C[-\pi,\pi]$, where
\begin{equation}\label{KAL}
\hat{K} A=\hat L_1 \ds\frac{A}{\left(\lambda_1-\lambda\right) \left(\lambda_2-\lambda\right)},
\end{equation}
is bounded uniformly in $\sigma$ for small $\sigma$. Here
$
R_0(t)=-D e^{\varepsilon k Y},
$
\begin{equation}\label{Dq}
D=- e^{-ka}P_0(k,\Lambda_1)\ds\frac{1}{\Lambda_2-\Lambda_1}\left(\ds\frac{k}{q_1\lambda'_1(k)}\right),
\end{equation}
$q_1$ in (\ref{Dq}) is defined in (\ref{pOG}) and $P_0$ in (\ref{P0}).
\end{lem}

Equation (\ref{All}) now can be rewritten as
\begin{equation}\label{aster}
A=\varepsilon\hat L_2\hat{N} \left(\hat{T}_0 A+\ds\frac{1}{\sigma} A(0)R_0(t) \right)
\end{equation}
or
\begin{equation*}
\left(1-\varepsilon\hat{T}\right)A=\ds\frac{\varepsilon}{\sigma} A(0)\hat L_2 \hat{N} R_0, \quad {\rm where}\quad\hat{T}=\hat L_2\hat{N}\hat{T}_0.
\end{equation*}
Thus we have
\begin{equation}\label{solution A}
A=\ds\frac{\varepsilon}{\sigma}A(0)\left(1-\varepsilon\hat{T}\right)^{-1} \hat L_2 \hat{N} R_0
\end{equation}
since $\hat{T}$ is bounded.\\
Substituting $p=0$, dividing by $A(0)$ and multiplying by $\sigma$, we obtain a secular equation for $\sigma$:
\begin{equation}\label{AsecL1}
\sigma=\varepsilon F\left(\varepsilon, \varepsilon_1, \sigma\right),
\end{equation}
where
$$
F\left(\varepsilon, \varepsilon_1, \sigma\right)=\left.\left(1-\varepsilon\hat{T}\right)^{-1}\hat L_2\hat{N} R_0\right\vert_{p=0}.
$$
In the leading term we obtain
$
\sigma=\varepsilon\left.\left(\hat L_2 \hat{N} R_0\right)\right\vert_{p=0}.
$
We have
\begin{equation*}
\left.L_2\right\vert_{p=0}=\dot{X} e^{-ak+\varepsilon k Y}=\dot{X} e^{-ak}\left(1+\varepsilon k Y+O(\varepsilon^2)\right),\quad \hat{N}=\hat{N}_0+O\left(\varepsilon^2\ln\varepsilon\right)
\end{equation*}
and hence, expanding $e^{\varepsilon kY}$ in $\varepsilon$ and using (\ref{lemma45}) and Proposition \ref{NN} as in section \ref{3.4}, we obtain
\begin{equation*}
\sigma
= \varepsilon^2 D e^{-ak} k\left(\frac{1}{2}S+\pi\mu\right)+O\left(\varepsilon^3\ln\varepsilon\right).
\end{equation*}
This proves Theorem\;\ref{LN4}. Note that, as in section \ref{3.4},  formula (\ref{solution A}) provides an exact solution of problem (\ref{etiqueta_1})-(\ref{etiqueta_2_4}), (\ref{2.b.b}).

\subsection{Complex resonances}\label{4.3}
In this section we prove Theorem \ref{LN5} for the resonance produced by the cylinder in the lower layer.
In this case we assume $\lambda=\Lambda_2(1-\sigma^2)=k(1-\sigma^2)$ and construct complex resonances lying close to the embedded threshold $\Lambda_2$.\\
As in section \ref{3.5}, the operator $\hat L_1$ possesses the same kernel $L_1(t,p)$, but  the integration is carried out along the contour $\gamma$ shown in Fig.\ref{X1}. The substitution (\ref{subst p}) holds as in section \ref{3.5}, and $A$ satisfies (\ref{All}) with the change of the contour of integration mentioned above. Similarly to Lemma\;\ref{Lamb_2}, we have (here $\hat K$ is still given by (\ref{KAL}) but with the contour change mentioned above)
\begin{lem}\label{YY}
The operator
\begin{equation*}
\hat{T}_0 A=\hat{K} A-\ds\frac{1}{\sigma}A(0)R_0(t)
\end{equation*}
acting from $\mathscr{A}$ to $C[-\pi,\pi]$, where
$
R_0(t)=-D e^{\varepsilon k Y},
$
\begin{equation}\label{Dq1}
D=e^{-ak}P_0(k,\Lambda_2) \ds\frac{k}{(\Lambda_2-\Lambda_1)q_2},
\end{equation}
and $q_2$ is defined in (\ref{p phiphi}), is bounded uniformly in $\sigma$ for small $\sigma$.
\end{lem}
Note that, although this lemma looks almost identical to Lemma \ref{XX_x}, the constants and contours are different. Equation (\ref{All}) for $A$, as above (cf. (\ref{aster})), can be rewritten as
\begin{equation*}
\begin{array}{lll}
A=\varepsilon \hat L_2\hat{N} \hat{K} A=\varepsilon \hat L_2\hat{N}\hat{T}_0 A+\ds\frac{\varepsilon}{\sigma}A(0)\hat L_2\hat{N} R_0,
\end{array}
\end{equation*}
and hence
\begin{equation*}
\left(1-\varepsilon\hat{T}\right) A=\ds\frac{\varepsilon}{\sigma}A(0) \hat L_2\hat{N} R_0,\quad {\rm where}\quad\hat{T}=\hat L_2 \hat{N}\hat{T}_0.
\end{equation*}
Thus
\begin{equation}\label{Allr}
A(p)=\ds\frac{\varepsilon}{\sigma}A(0)\left(1-\varepsilon\hat{T}\right)^{-1}\hat L_2\hat{N} R_0.
\end{equation}
As in sections\;3.5-3.6, if
\begin{equation}\label{OCS}
A(\pm p_1)=0
\end{equation}
(this is the analogue of the orthogonality conditions (\ref{orc})), then (\ref{Allr}) defines a trapped mode. It turns out that the orthogonality conditions (\ref{OCS}) cannot be satisfied for  problem $L$ in contrast to problem $U$. Let us verify this statement.
In the leading term, up to a nonzero factor, (\ref{OCS}) reads
$
\left.\mathcal{F}(p)\right\vert_{\pm p_1}:=\left.\hat L_2 \hat{N} R_0\right\vert_{\pm p_1}=0.
$
Again, up to a nonzero factor, we have
\begin{equation*}
\mathcal{F}(p)=\ds\int\limits_{-\pi}^{\pi} \left(\ds\frac{ip}{\tau}\dot{Y}+\dot{X}\right)  e^{-(a-\varepsilon Y)\tau-ip\varepsilon X} \hat{N} e^{\varepsilon k Y}\,dt.
\end{equation*}
Note that the last expression at the point $p=-p_1$ is simply the complex conjugate of it at $p=p_1$. Thus  we have only to calculate its value at $p=p_1$.

We have, performing  the calculations similarly to section\;3.5,
\begin{align*}
\Re\mathcal{F}(p_1)&=  -\varepsilon e^{-a \tau(p_1)}\left[\ds\frac{1}{2}\;\ds\frac{k^2}{\tau(p_1)} S+\pi k\mu\right]+ O\left(\varepsilon^2\ln\varepsilon\right)\neq0,\\\\
\Im\mathcal{F}(p_1)&= \varepsilon e^{-a \tau(p_1)}\;\ds\frac{p_1k}{\tau(p_1)}\;\pi\nu  +O\left(\varepsilon^2\ln\varepsilon\right).
\end{align*}
Thus we see that (\ref{OCS}) cannot be true for sufficiently small $\varepsilon$.\\

We complete the proof of Theorem \ref{LN5}. Let us calculate the leading terms of the real and imaginary parts of $\sigma$, which satisfies, just as in the preceding section, the secular equation
\begin{equation}\label{sec eq}
\sigma=\varepsilon F\left(\varepsilon,\varepsilon_1, \sigma\right),
\end{equation}
where
$$
F\left(\varepsilon,\varepsilon_1, \sigma\right)=\left.\left(\left(1-\varepsilon\hat{T}\right)^{-1}\;\hat L_2 \hat{N} R_0\right)\right\vert_{p=0}.
$$
In the leading term we have, just as in the preceding subsection,
$$
\sigma=\varepsilon \left.\hat{L}_2 \hat{N} R_0\right\vert_{p=0}+O\left(\varepsilon^3\right)
$$
and we see that
$$
\sigma=\frac{\varepsilon^2}2De^{-ak}(S+2\pi\mu)+O(\varepsilon^3\ln\varepsilon)
 $$
 with the constant $D$ given by (\ref{Dq1}) and hence we have proven (\ref{re-s-L}).\\
Let us calculate the imaginary part of $\sigma$. The corresponding computation is very similar to section\;3.5. The principal contribution to $\Im \sigma$ comes from the term
\begin{equation}\label{termep}
\varepsilon^2\Im\left(\left.\hat{T} \hat L_2 \hat{N} R_0\right\vert_{p=0}\right),
\end{equation}
which, in its turn, comes from the half-residues of the expression $\hat{T}_0 \hat L_2 \hat{N} R_0$. Similarly to section\;3.5, let
\begin{equation}\label{f}
f=f_1+if_2=\hat L_2 \hat{N} R_0.
\end{equation}
Clearly, $\hat{N} R_0$ is purely real and hence
\begin{equation}\label{f12}
f_1=(\Re \hat L_2)\hat{N} R_0 ,\qquad f_2=(\Im \hat L_2 )\hat{N} R_0,
\end{equation}
where $\Re \hat{L}_2$ and $\Im\hat{L}_2$ mean the integral operators with the kernels $\Re L_2$ and $\Im L_2$, respectively.
Consider the expression
\begin{equation*}
\hat{K} f= \ds\int\limits_{\gamma} L_1\ds\frac{f}{(\lambda_1-\lambda)(\lambda_2-\lambda)}dp
\end{equation*}
which enters (\ref{termep}) through $\hat{T}_0$.
By an argument similar to Lemma\;\ref{XX}, we have for $f$ given by (\ref{f})
\begin{equation*}
\hat{K} f= \ds\fint L_1\;\ds\frac{f}{(\lambda_1-\lambda)(\lambda_2-\lambda)}dp+2\pi i\left.\left(\Res_{p_1}\left(\ds\frac{L_1 f}{(\lambda_1-\lambda)(\lambda_2-\lambda)}\right)\right)\right|_{\sigma=0}+O(\sigma^2).
\end{equation*}
We see that
\begin{equation}\label{Im sec}
\Im \hat{K} f=2\pi\left.\left(\ds\frac{(L_{11} f_1-L_{12}f_2)\tau(p_1)}{\lambda_1'(\tau(p_1))p_1(\lambda_2-\lambda)}\right)\right|_{\sigma=0}+O(\sigma^2),
\end{equation}
where $L_{11}=\Re L_1$, $L_{12}=\Im L_1$, that is,
\begin{equation*}
L_{11}=\ds\frac{1}{2\pi} P_0(\tau,\lambda) e^{-(a-\varepsilon Y)\tau} \cos\varepsilon p X,\quad
L_{12}=\ds\frac{1}{2\pi} P_0(\tau,\lambda) e^{-(a-\varepsilon Y)\tau} \sin\varepsilon p X,
\end{equation*}
and, by (\ref{f12}),
\begin{align*}
f_1&=-\ds\int\limits_{-\pi}^{\pi}\left(\dot{X}\cos \varepsilon p X+\ds\frac{p\dot{Y}}{\tau} \sin\varepsilon p X\right) e^{-a\tau+\varepsilon \tau Y}\;\hat{N} D e^{\varepsilon k Y}\;dt,\\\\
f_2&=-\ds\int\limits_{-\pi}^{\pi}\left(\ds\frac{p\dot{Y}}{\tau} \cos\varepsilon p X-\dot{X}\sin \varepsilon p X\right) e^{-a\tau+\varepsilon \tau Y}\;\hat{N} D e^{\varepsilon k Y}\;dt.
\end{align*}
Let us calculate the asymptotics of $f_{1,2}$. We have, similarly to section\;3.5,
\begin{equation*}
f_1= \varepsilon D e^{-a\tau}\left(\ds\frac{1}{2}\;\ds\frac{k^2}{\tau} S+\pi k\mu\right)+O\left(\varepsilon^2\ln\varepsilon\right),\quad
f_2= \varepsilon D e^{-a\tau}\left(-\ds\frac{p k}{\tau}\;\pi\nu\right)+O\Big(\varepsilon^2\ln\varepsilon\Big).
\end{equation*}
Substituting in (\ref{Im sec}), we obtain
\begin{equation*}
\Im \hat{K}f =-\varepsilon DD_1 e^{-2a\tau_1+\varepsilon\tau_1 Y} \left[\left(\ds\frac{1}{2}\;\ds\frac{k^2}{\tau_1} S+\pi k\mu\right)\;\cos\varepsilon p_1^0 X+
\ds\frac{p_1^0 k}{\tau_1}\;\pi\nu \sin\varepsilon p_1^0 X\right]+O\left(\varepsilon^2\ln\varepsilon+\varepsilon\sigma^2\right),
\end{equation*}
where
\begin{equation}\label{gammaL}
D_1=-\frac{P_0(\tau_1,\Lambda_2)\tau_1}{(\tau_1-k)\lambda_1'(\tau_1)p_1^0}
\end{equation}
and we recall that $\tau_1$ is the same as in (\ref{mathcal{R}}), $\tau(p_1^0)=\tau_1$.
Finally, substituting in (\ref{termep}) we  obtain for the solution $\sigma$ of (\ref{sec eq})
\begin{align}\nonumber
\Im \sigma&=-\varepsilon^3 DD_1 e^{-2a\tau_1-ka}\ds\int\limits_{-\pi}^{\pi}\dot{X} e^{\varepsilon k Y}\hat{N} e^{\varepsilon\tau_1 Y}\Bigg[\left(\frac{1}{2}\frac{k^2}{\tau_1} S+\pi k\mu\right) \cos\varepsilon p_1^0 X\\\label{ImsigL}
&\qquad\qquad\qquad\qquad\qquad+\ds\frac{p_1^0 k}{\tau_1} \pi\nu\sin\varepsilon p_1^0 X\Bigg]\,dt+O\left(\varepsilon^5\ln\varepsilon\right).
\end{align}
We have, expanding in $\ve$, using (\ref{lemma45}) and Proposition\;\ref{NN},
\begin{align*}
&\ds\int\limits_{-\pi}^{\pi}\dot{X} e^{\varepsilon k Y}\hat{N} e^{\varepsilon\tau_1 Y}\left[\left(\frac{1}{2}\frac{k^2}{\tau_1} S+\pi k\mu\right) \cos \varepsilon p_1^0 X+
\frac{p_1^0 k}{\tau_1} \pi\nu\sin \varepsilon p_1^0 X \right]\,dt\\
&\qquad\qquad=-\varepsilon\left(\ds\frac{(p_1^0)^2 k}{\tau_1} (\pi\nu)^2+\ds\frac{k}{\tau_1}\left(\ds\frac{1}{2}\;kS+\pi\tau_1\mu\right)^2\right)+O\left(\varepsilon^2\ln\varepsilon\right).
\end{align*}
Substituting in (\ref{ImsigL}), we obtain (\ref{im-s-L}), and hence we have proven Theorem\;\ref{LN5}.\qquad$\blacksquare$

\appendix

\section{Appendix\;1. Integral equations for the cylinder in the upper layer}
\setcounter{equation}{0}

\subsection{Integral equations for $\varphi, \psi, \theta$}

Since $\phi_1$ satisfies  equation (\ref{etiqueta_1}), we have
by the Green formula,
\begin{equation}\label{GF}
\phi_{1}(\xi,\eta)=-\int\limits_{\Gamma_{F}+\Gamma_{I}+\Gamma} G(x-\xi, y-\eta)\frac{\partial\phi_1(x,y)}{\partial n}\,dl+
\int\limits_{\Gamma_{F}+\Gamma_{I}+\Gamma} \frac{\partial G(x-\xi, y-\eta)}{\partial n}\phi_1(x,y)\,dl,
\end{equation}
where $G(x,y)=-\frac{1}{2\pi} K_0\left(k\sqrt{x^2+y^2}\right)$
is the fundamental solution of (\ref{etiqueta_1}), the normal derivatives are taken at the point $(x,y)$ and $dl$ is the arc element at the same point. In order to obtain the equations for $\varphi, \psi, \theta$, we will perform in (\ref{GF}) three passages to the limit as
$$
(\xi,\eta)\to \Gamma_{F},\quad (\xi,\eta) \to \Gamma_{I},\quad
(\xi,\eta)\to \Gamma.
$$
We have for the normals on $\Gamma, \Gamma_{F}, \Gamma_{I}$ (see (\ref{bfmr}) for the definition of ${\bf m}$)
$$
{\rm on}\;\Gamma:\quad{\bf n}=\displaystyle\frac{\bf m}{|{\bf m}|},\qquad
{\rm on}\;\Gamma_{F}:\quad{\bf n}=(0,1),\qquad {\rm on}\;\Gamma_{I}:\quad{\bf n}=(0,-1).
$$
Using the jump conditions (see, e.g.,\cite{Kuz-Maz-Va}) for the potentials in (\ref{GF}) and equations (\ref{etiqueta_2_2}), (\ref{Def}), we obtain on $\Gamma_F$
\begin{align}\nonumber
\frac{1}{2}\varphi(\xi)&= -\lambda\int\limits_{\Gamma_{F}} G\varphi\,dx+\int\limits_{\Gamma_{I}} G\psi\,dx+\int\limits_{\Gamma_{F}} (\nabla G\cdot {\bf n})\varphi\,dx\\\label{GaF}
&\quad+ \int\limits_{\Gamma_{I}} (\nabla G\cdot {\bf n}) \phi_1(x,-b)\,dx+\int\limits_{\Gamma} (\nabla G\cdot {\bf n}) \theta dl,
\end{align}
where the arguments of $G$ and $\nabla G$ are taken according to formula (\ref{GF}) with $(\xi,\eta)=(\xi,0)\in\Gamma_{F}$ and $(x,y)$ belonging to $\Gamma_{F}$, $\Gamma_{I}$ or $\Gamma$, respectively. We have $\nabla G\cdot {\bf n}=0$ on $\Gamma_{F} $ and
\begin{equation*}
\nabla G\cdot {\bf n}=-\frac{1}{2\pi}K'_0 \left(k d_{F I}(x,\xi)\right) \frac{kb}{d_{F I}(x,\xi)}\quad {\rm on }\quad\Gamma_{I};
\end{equation*}
here
$
d_{FI}(x,\xi)=|{\bf r}_{FI}(x,\xi)|$, ${\bf r}_{F I}(x,\xi)=(x-\xi, -b).
$
Thus (\ref{GaF}) takes the form
\begin{align}
\nonumber\varphi(\xi)&=\frac{1}{\pi} \lambda\int K_0\left(k|x-\xi|\right)\varphi(x)\,dx-\frac{1}{\pi} \int K_0\left(k d_{F I}(x,\xi)\right)\psi(x)\,dx\\\label{gamma f}
&\quad-\frac{kb}{\pi}\int\frac{K'_{0}\left(k d_{F I}(x,\xi)\right)}{d_{F I}(x,\xi)}\phi_{1}(x,-b)\,dx-\frac{\varepsilon k}{\pi}\int\limits_{-\pi}^{\pi}{\bf r}_{F \Gamma}(\xi,t)\cdot{\bf m}(t)
\frac{K'_0\left(kd_{F \Gamma}(\xi,t)\right)}{d_{F \Gamma}(\xi,t)}\theta(t)\,dt,
\end{align}
where
\begin{equation}\label{d_1}
{\bf r}_{F \Gamma}(\xi,t)=(\varepsilon X(t)-\xi,\varepsilon Y(t)-a),\qquad d_{F\Gamma}(\xi,t)=|{\bf r}_{F\Gamma}(\xi,t)|.
\end{equation}
Likewise, on $\Gamma_{I}$ we obtain
\begin{equation*}
\frac{1}{2}\phi_1(\xi,-b)=-\lambda \int\limits_{\Gamma_{F}} G\varphi\,dx+\int\limits_{\Gamma_{I}} G\psi\,dx+\int\limits_{\Gamma_{F}}(\nabla G\cdot {\bf n})\varphi \,dx+\int\limits_{\Gamma_{I}}(\nabla G\cdot {\bf n})\phi_1(x,-b)\, dx+\int\limits_{\Gamma}(\nabla G\cdot {\bf n})\theta\, dl.
\end{equation*}
Noting that $\nabla G\cdot {\bf n}=0$ on $\Gamma_{I}$, we obtain
\begin{align}\nonumber
\phi_1(\xi,-b)&=\frac{\lambda}{\pi}\ds\int K_0\left(k d_{FI}(x,\xi)\right) \varphi(x)\,dx-\frac{1}{\pi}\int K_0\left(k|x-\xi|\right)\psi(x)\,dx\\\label{gamma i}
&\quad-\frac{kb}{\pi}\int\frac{K'_0\left(k d_{FI}(x,\xi)\right)}{d_{FI}(x,\xi)}\varphi(x)\,dx-\frac{\varepsilon k}{\pi}\int\limits_{-\pi}^{\pi}{\bf r}_{I\Gamma}(\xi,t)\cdot{\bf m}(t)\frac{K'_0\left(kd_{I\Gamma}(\xi,t)\right)}{ d_{I\Gamma}(\xi,t)}\theta(t)\,dt,
\end{align}
\begin{equation}\label{d_2}
{\bf r}_{I\Gamma}(\xi,t)=(\varepsilon X(t)-\xi,\varepsilon Y(t)+c), \quad d_{I\Gamma}(\xi,t)=|{\bf r}_{I\Gamma}(\xi,t)|,\quad c=b-a.
\end{equation}
Similarly, on $\Gamma$ we have
\begin{equation*}
\ds\frac{1}{2}\theta=-\lambda \int\limits_{\Gamma_{F}} G\varphi\,dx+\int\limits_{\Gamma_{I}} G\psi\,dx+\int\limits_{\Gamma_{F}}(\nabla G\cdot {\bf n})\varphi\, dx+\int\limits_{\Gamma_{I}}(\nabla G\cdot {\bf n})\phi_1(x,-b)\, dx+\int\limits_{\Gamma}(\nabla G\cdot {\bf n})\theta\, dl.
\end{equation*}
Denote
\begin{equation}\label{{d_3}}
d(t,s)=|{\bf r}(t)-{\bf r}(s)|,
\end{equation}
here ${\bf r}$ is defined in (\ref{bfmr}).
In this notation, the last equation takes the form
\begin{align}\nonumber
 \theta(t)&=\frac{1}{\pi}\int K_0\left(k d_{F\Gamma }(x,t)\right)\lambda\varphi(x)\,dx-
 \frac{1}{\pi}\int K_0\left(k d_{I\Gamma}(x,t)\right)\psi(x)\,dx\\\nonumber
&\quad-\frac{1}{\pi}\int kK'_0\left(k d_{F\Gamma }(x,t)\right)\frac{a-\varepsilon Y}{d_{F\Gamma}(x,t)}\varphi(x)\,dx-
\frac{1}{\pi}\int kK'_0\left(k d_{I\Gamma}(x,t)\right)\frac{c+\varepsilon Y(t)}{d_{I\Gamma}(x,t)}\phi_1(x,-b)\,dx\\\label{gamma}
&\quad-\frac{\varepsilon k}{\pi}\int\limits_{-\pi}^{\pi} \frac{K'_0\left(\varepsilon k d(t,s)\right)}{d(t,s)}
 ({\bf r}(s)-{\bf r}(t))\cdot
 {\bf m}(s)\theta(s)\,ds.
\end{align}
Since $\left.\phi_1\right|_{y=-b}$ is expressed through $\psi$ by (\ref{phi1 psi}), equations (\ref{gamma f}), (\ref{gamma i}), (\ref{gamma}) represent a system for $\varphi$, $\psi$, $\theta$.

\subsection{Fourier transform of the integral equations for $\varphi$, $\psi$, $\theta$}

Here we will reduce the obtained integral equations  (\ref{gamma f}), (\ref{gamma i}), (\ref{gamma}) to three integral equations for the Fourier transforms $\tilde{\varphi}$, $\tilde{\psi}$ and $\theta$.
We will need the following formulas:
\begin{align}\label{ft1}
F_{\xi\to p}\left[K_0(k |\xi|)\right]&={\pi}{\tau^{-1}},
\\\label{ft2}
F_{\xi\to p}\left[K_0\left(k\rho\right)\right]&= {\pi}{\tau}^{-1}e^{-h\tau},
\\\label{ft3}
F_{\xi\to p}\left[{k\rho^{-1}\xi K'_0(k\rho)}\right]&= {i\pi p}{\tau}^{-1}e^{-h\tau},
\\\label{ft4}
F_{\xi\to p}\left[{k\rho^{-1} K'_0\left(k\rho\right)}\right]&= -{\pi}{h}^{-1}e^{-h\tau},
\end{align}
where $\tau=\sqrt{k^2+p^2}$, $\rho=\sqrt{\xi^2+h^2}$, and $h>0$ is a parameter (see \cite{Grad}, formulas (6.672.14) and (6.677.5)).

Applying the Fourier transform (\ref{Fourier}) to (\ref{gamma f}), using the formula for the Fourier transform of a convolution, formula (\ref{ft1}) for the first term in the right-hand side, formula (\ref{ft2}) for the second term, and formula (\ref{ft4}) for the third term, we come to
\begin{align*}
\tilde{\varphi}(p)&=\frac{\lambda\tilde{\varphi}(p)}{\tau}-\frac{e^{-b\tau}}{\tau}\tilde{\psi}(p)+ e^{-b\tau}\tilde{\phi}_1(p,-b)\\
&\quad+\frac{\varepsilon k}{\pi}\int\limits_{-\pi}^{\pi} F_{x\to p}\left[ -{\bf r}_{F\Gamma}(x,t)\cdot{\bf m}(t) \frac{K'_0\left(kd_{F\Gamma}(x,t)\right)}{d_{\Gamma F}(x,t)}\right]\theta(t)\,dt.
\end{align*}
By (\ref{ft3}), (\ref{ft4}) and (\ref{d_1}) we have, changing $x-\varepsilon X$ to $\xi$ and with $h=a-\varepsilon Y$,
$$
F_{x\to p} \left[\frac{k(x-\varepsilon X) K'_0\left(k d_{F\Gamma}(x,t)\right)}{d_{F\Gamma}(x,t)}\right]=\frac{i\pi p}{\tau}\;e^{-(a-\varepsilon Y)\tau-ip\varepsilon X}
$$

$$
F_{x\to p} \left[\frac{k(a-\varepsilon Y)K'_0\left(kd_{F\Gamma}(x,t)\right)}{d_{F\Gamma}(x,t)}\right]=-\pi e^{-(a-\varepsilon Y)\tau-ip\varepsilon X}.
$$
Finally, using the last two formulas and (\ref{phi1 psi}) we come to (\ref{M4})
instead of (\ref{gamma f}).\\
Likewise, applying the Fourier transform to (\ref{gamma i}) and using (\ref{d_2}) and (\ref{phi1 psi}), we obtain (\ref{M5}) in a similar way instead of (\ref{gamma i}).

Equation (\ref{gamma}) has the form
\begin{align}\nonumber
\theta+\hat{M}_1\theta&=\frac{1}{\pi}\int K_0\left(k d_{F\Gamma}(x,t)\right)\lambda\varphi(x)\,dx
-\frac{1}{\pi}\int K_0\left(k d_{I\Gamma}(x,t)\right)\psi(x)\,dx\\\nonumber
&\quad-\frac{1}{\pi}\int k K'_0\left(k d_{F\Gamma}(x,t)\right)\frac{a-\varepsilon Y}{d_{F\Gamma }(x,t)}\varphi(x)\,dx-\\\label{eqth}
&\quad-\frac{1}{\pi}\int k K'_0\left(k d_{I\Gamma}(x,t)\right)\frac{(c+\varepsilon Y)}{d_{I\Gamma}(x,t)} \phi_1(x,-b)\,dx,
\end{align}
where $\hat M_1$ is given by (\ref{M_1the}), (\ref{M_1'}).
Let us express the right-hand side of (\ref{eqth}) in terms of the Fourier transforms of $\varphi$ and $\psi$. For the first summand we have, similarly to the above, and using (\ref{ft2}),
\begin{equation*}
\frac{1}{\pi}\int K_0\left(k d_{F\Gamma}(x,t)\right)\lambda\frac{1}{2\pi}\int e^{ipx}\tilde{\varphi}(p)\,dp dx
=\frac{\lambda}{2\pi}\int\frac1\tau{e^{-(a-\varepsilon Y)\tau+ip\varepsilon X}}\tilde{\varphi}(p)\,dp.
\end{equation*}
For the second summand, we have similarly
$$
\frac{1}{\pi} \int K_0\left(k d_{I\Gamma}(x,t)\right)\psi(x)\,dx
=\frac1{2\pi}\int\frac1\tau{e^{-(c+\varepsilon Y)\tau+ip\varepsilon X}}\tilde{\psi}(p)\,dp.
$$
For the third summand, we have, using (\ref{ft4})
$$
\frac{1}{\pi} \int k K'_0\left(k d_{F\Gamma}(x,t)\right)
\frac{a-\varepsilon Y}{d_{F\Gamma}(x,t)}\varphi(x)\,dx
=-\frac{1}{2\pi}\int e^{-(a-\varepsilon Y)\tau+ip\varepsilon X}\tilde{\varphi}(p)\,dp.
$$
For the last summand, we have similarly
$$
\frac{1}{\pi}\int k K'_0\left(k d_{I\Gamma}(x,t)\right)
\frac{c+\varepsilon Y}{d_{I\Gamma}(x,t)}\phi_1(x,-b)\,dx
=-\frac{1}{2\pi}\int e^{-(c+\varepsilon Y)\tau+ip\varepsilon X}\cdot \tilde{\phi}_1(p,-b)\,dp.
$$
Finally, (\ref{eqth}) takes the form (\ref{M1}). Thus we have obtained equations (\ref{M4})-(\ref{M1}).

\section{Appendix\;2. Integral equations for the cylinder in the lower layer}
\setcounter{equation}{0}

\subsection{Integral equations for $\varphi$, $\theta$}

Since $\phi_2$ satisfies (\ref{etiqueta_1_2}) in $\Omega_2$, we have, by the Green formula
\begin{equation}\label{GF1}
\phi_2(\xi,\eta)=-\int\limits_{\Gamma_{I}+\Gamma} G\left(x-\xi, y-\eta\right)\frac{\partial\phi_2(x,y)}{\partial n}\,dl+\int\limits_{\Gamma_{I}+\Gamma} \frac{\partial G(x-\xi, y-\eta)}{\partial n}\phi_2(x,y)\,dl.
\end{equation}
Denote
$\psi:=\left.\phi_{1y}\right\vert_{y=-b}=\left.\phi_{2y}\right\vert_{y=-b}$. Passing to the limits $(\xi,\eta)\to \Gamma_{I}$ and $(\xi,\eta)\to \Gamma$, we obtain similarly to Appendix\;1,
\begin{equation}\label{Gammtheta}
\phi_2(\xi,-b)= \frac{1}{\pi}\int K_0\left(k|x-\xi|\right)\psi(x)\,dx
-\frac{\varepsilon k}{\pi} \int\limits_{-\pi}^{\pi}\frac{K'_0\left(kd_{I\Gamma}(\xi, t)\right)}{d_{I\Gamma}(\xi, t)}{\bf r}_{I\Gamma}(\xi,t)\cdot {\bf m}(t) \theta(t)\,dt,
\end{equation}
\begin{align}\nonumber
\theta(t)&= \frac{1}{\pi}\int K_0\left(kd_{I\Gamma}(x, t)\right)\psi(x)\,dx-\frac{1}{\pi}\ds\int\limits_{-\infty}^{\infty} \frac{kK'_0\left(kd_{I\Gamma}(x, t)\right)}{d_{I\Gamma}(x, t)}({a-\varepsilon Y})\phi_2(x,-b)\,dx\\\label{Gammtheta1}
&\quad-\frac{\varepsilon k}{\pi} \ds\int\limits_{-\pi}^{\pi} \ds\frac{K'_0\left(\varepsilon k d(s,t)\right)}{d(s,t)}\left({\bf r}(s)-{\bf r}(t)\right)\cdot{\bf m}(s) \theta(s)\,ds,
\end{align}
where $d(s,t)$ is defined in (\ref{{d_3}}) and
$
{\bf r}_{I\Gamma}(\xi,t)=(\varepsilon X(t)-\xi,\varepsilon Y(t)-a)$, $d_{I\Gamma}(\xi,t)=|{\bf r}_{I\Gamma}(\xi,t)|.
$
Since $\psi(x)$ and $\phi_2(x,-b)$ are expressed through $\varphi$ by (\ref{psi1}), (\ref{Phi2}),  equations (\ref{Gammtheta}) and (\ref{Gammtheta1}) represent a system for $\varphi, \theta$.

\subsection{Fourier transform of the integral equations for $\varphi$, $\theta$}

Similarly to Appendix\;1, using formulas (\ref{ft1})-(\ref{ft4}) we obtain for the Fourier transforms of $\phi_2(x,-b)$, $\psi(x)$ and $\theta(t)$ the following system:
\begin{equation}\label{tilde{phi}_2}
\tilde{\phi}_2(p,-b)=\ds\frac{1}{\tau} \tilde{\psi}(p)-\varepsilon\ds\int\limits_{-\pi}^{\pi} \left(\frac{ip}{\tau}\dot{Y}+\dot{X}\right) e^{-(a-\varepsilon Y)\tau-ip\varepsilon X}\theta(t)\,dt
\end{equation}
\begin{equation}\label{hat{M}}
(1+\hat{M}_1)\theta=\frac{1}{2\pi}\ds\int\ds\frac{1}{\tau}  e^{-(a-\varepsilon Y)\tau+ip\varepsilon X}\tilde{\psi}(p)\,dp+\ds\frac{1}{2\pi}\ds\int  e^{-(a-\varepsilon Y)\tau+ip\varepsilon X}\tilde{\phi}_2(p,-b)\,dp.
\end{equation}

\section{Appendix\;3. Integral formulas for the dipole strengths.}
\setcounter{equation}{0}
In this appendix we collect some formulas needed in the derivation of the asymptotics of the eigenvalues and resonances. First of all, we note an obvious formula for the operator $\hat{M}_1^{(0)}$ from (\ref{M_1}) (see (\ref{M_1^{0}})) for a detailed notation):
\begin{equation}\label{def M1phi}
\hat{M}_1^{(0)} f=-2\ds\int\limits_{C} \ds\frac{\partial G_0}{\partial n}\;f\;dl,
\end{equation}
$G_0(x,y)=\ds\frac{1}{2\pi}\ln\sqrt{x^2+y^2}$. In this notation, the Gauss law reads
$
\hat{M}_1^{(0)}1=1
$
and hence
\begin{equation}\label{lemma45}
\left(1+\hat{M}_1^{(0)}\right)^{-1}\;1=\ds\frac{1}{2}.
\end{equation}
\begin{lem}\label{DN}
Let $\Phi_{i}$ and $\Phi_{e}$ satisfy the Laplace equation in the interior and the exterior of $C$, respectively, $|\nabla\Phi_e|\to 0$ as $x,y\to \infty$. Moreover, let
$$
\left.\Phi_{i}\right\vert_{C}=p,\quad \left.\Phi_{e}\right\vert_{C}=q,\quad \left.\frac{\partial\Phi_{i}}{\partial n}\right\vert_{C}=\left.\frac{\partial\Phi_{e}}{\partial n}\right\vert_{C}=f,
$$
where $p, q, f$ are sufficiently smooth and $\ds\int\limits_{C} f\;dl=0$. Then
\begin{equation}\label{p-q}
2\left(1+\hat{M}_1^{(0)}\right)^{-1} p=p-q,
\end{equation}
where $M_1^{(0)}$ is given by (\ref{def M1phi}).
\end{lem}

\no {\bf Proof}. Equality (\ref{p-q}) means that
$
2p=\left(1+\hat{M}_1^{(0)}\right) p-\left(1+\hat{M}_1^{(0)}\right) q
$
or
\begin{equation}\label{p+q}
p+q+\hat{M}_1^{(0)}(q-p)=0.
\end{equation}
But, by the Green formula and the jump conditions, $p$ and $q$ satisfy
\begin{equation}\label{p}
\left(1-\hat{M}_1^{(0)}\right) p= 2\ds\int\limits_{C} G_0 f\,dl,\quad
\left(1+\hat{M}_1^{(0)}\right) q= -2\ds\int\limits_{C} G_0 f\,dl.
\end{equation}
Summing the two equations in (\ref{p}), we obtain (\ref{p+q}).$\qquad\blacksquare$


\begin{prop}\label{NN}
The following formulas are valid:
\begin{equation}\label{mu1}
\int\limits_{-\pi}^{\pi}\dot{X}(t)\hat{N}_0\; Y(t)\,dt=-\pi\mu,\quad \int\limits_{-\pi}^{\pi}\dot{Y}(t)\hat{N}_0 X(t)\, dt=\pi\kappa,
\end{equation}
\begin{equation}
 \label{nu1}
\int\limits_{-\pi}^{\pi}\dot{X}(t)\hat{N}_0 X(t)\,dt=-\int\limits_{-\pi}^{\pi}\dot{Y}(t)\hat{N}_0 Y(t)\,dt=-\pi\nu,
\end{equation}
where $\hat{N}_0=\left(1+\hat{M}_1^{(0)}\right)^{-1}$ and $\kappa$ is defined by (\ref{kappa}) below (we do not use the  formula for $\kappa$ and present it for completeness).
\end{prop}
{\bf Proof}.
Let $\Phi_{1,2}$ be solutions of the following problems:
$$
\Delta\Phi_{1,2}=0~{\rm in}~\Omega_0,\quad \left.\frac{\partial \Phi_{1,2}}{\partial n}\right\vert_{C}=n_{1,2},\quad \nabla\Phi_{1,2}\to0,\quad r\to\infty.
$$
Note that $\Phi_2=\Psi$, the solution of (\ref{NP}). It is known that \cite{Newman}
\begin{equation*}
\Phi_1=\const-\kappa\ds\frac{x}{r^2}-\nu\ds\frac{y}{r^2}+O\left(\ds\frac{1}{r^2}\right),\quad
\Phi_2=\const-\mu\ds\frac{y}{r^2}-\nu\ds\frac{x}{r^2}+O\left(\ds\frac{1}{r^2}\right).
\end{equation*}
It is also known that
\begin{equation}\label{kappa}
\kappa=\frac{1}{2\pi}\left(S+\int\limits_{C} \Phi_1\frac{\partial \Phi_1}{\partial n}\,dl\right),
 \quad
\mu=\frac{1}{2\pi}\left(S+\int\limits_{C} \Phi_2\frac{\partial \Phi_2}{\partial n}\,dl\right),
\end{equation}\begin{equation}\label{nu}
\nu=\frac{1}{2\pi}\int\limits_{C} \Phi_1\frac{\partial \Phi_2}{\partial n}\,dl=\frac{1}{2\pi}\int\limits_{C} \Phi_2\frac{\partial \Phi_1}{\partial n}\,dl.
\end{equation}
By Lemma\;\ref{DN} with $\Phi_{i}=y,~~\Phi_{e}=\Phi_2=\Psi$, we have
\begin{equation}\label{NPhiY}
\hat{N}_0 Y=\frac{1}{2}\left(Y-\left.\Phi_2\right\vert_{C}\right)
.\end{equation}
Substituting in (\ref{mu1}), we have  by (\ref{kappa})
$$
\int\limits_{-\pi}^{\pi} \dot{X} \hat{N}_0 Y\,dt=\frac{1}{2}\int\limits_{-\pi}^{\pi}\dot{X} \left(Y-\left.\Phi_2\right\vert_{C}\right)\,dt= -\frac{1}{2} S-\frac{1}{2}\int\limits_{C} n_2 \left.\Phi_2\right\vert_{C}\,dl=-\pi\mu,
$$
and hence the first formula from (\ref{mu1}) is proven.
Substituting (\ref{NPhiY}) in (\ref{nu1}), we have by (\ref{nu})
$$
\int\limits_{-\pi}^{\pi} \dot{Y} \hat{N}_0 Y\,dt=\frac{1}{2}\int\limits_{-\pi}^{\pi} \dot{Y} \left(Y-\left.\Phi_2\right\vert_{C}\right)\,dt= \frac{1}{2} \int\limits_{C} n_1 \left.\Phi_2\right\vert_{C}\,dl=\pi\nu.
$$
Likewise, by Lemma\;\ref{DN} with $\Phi_{i}=x$, $\Phi_{e}=\Phi_{1}$, we have
$
\hat{N}_0 X=\frac{1}{2}\left(X-\left.\Phi_1\right\vert_{C}\right)
$.
Substituting in (\ref{nu1}), we obtain by (\ref{nu})
$$
\int\limits_{-\pi}^{\pi} \dot{X} \hat{N}_0 X\,dt=\frac{1}{2}\int\limits_{-\pi}^{\pi}\dot{X} \left(X-\left.\Phi_1\right\vert_{C}\right)\,dt= -\frac{1}{2}\int\limits_{C} n_2\left. \Phi_1\right\vert_{C}\,dl=-\pi\nu,
$$
and hence formula (\ref{nu1}) is proven.
Finally, by (\ref{kappa}),
$$
\int\limits_{-\pi}^{\pi} \dot{Y} \hat{N}_0 X\,dt=\frac{1}{2}\int\limits_{-\pi}^{\pi}\dot{Y} \left(X-\left.\Phi_1\right\vert_{C}\right)\,dt= \frac12 S+\frac{1}{2}\int\limits_{C} n_1\left. \Phi_1\right\vert_{C}\,dl=\pi\kappa,
$$
and hence the second formula from (\ref{mu1}) is also proven. $\qquad\blacksquare$


\section{Acknowledgements}
PZ and  AM are grateful to CONACYT-M\'exico and CIC-UMSNH,  JEDM is grateful to CONACYT-M\'exico,   and MIRR is grateful to Vicerrector\'\i a de  Investigaci\'on de la UMNG  for partial financial support.

\end{document}